\documentclass{article}
\usepackage[hidelinks]{hyperref}
\usepackage[links]{agda}

\usepackage{amssymb}
\usepackage{amsmath}
\usepackage[utf8]{inputenc}
\usepackage[T1]{fontenc}
\usepackage{alphabeta}

\usepackage{bbm}
\usepackage{mathrsfs}
\usepackage{attrib}

\usepackage{newunicodechar}
\newunicodechar{⊗}{\ensuremath{\otimes}}
\newunicodechar{⊕}{\ensuremath{\oplus}}
\newunicodechar{𝕋}{\ensuremath{\mathbb{T}}}
\newunicodechar{⇒}{\ensuremath{\Rightarrow}}
\newunicodechar{⟶}{\ensuremath{\to}}
\newunicodechar{∋}{\ensuremath{\ni}}
\newunicodechar{◂}{\ensuremath{\mathbin{\triangleleft}}}
\newunicodechar{≪}{\ensuremath{\mathbin{\text{\guillemotleft}}}}
\newunicodechar{∅}{\ensuremath{\varnothing}}
\newunicodechar{𝕫}{\ensuremath{\mathsf{z}}}
\newunicodechar{𝕤}{\ensuremath{\mathsf{s}}}
\newunicodechar{𝕧}{\ensuremath{\mathsf{v}}}
\newunicodechar{∀}{\ensuremath{\mathnormal\forall}}
\newunicodechar{⊢}{\ensuremath{\vdash}}
\newunicodechar{₀}{\ensuremath{_0}}
\newunicodechar{₁}{\ensuremath{_1}}
\newunicodechar{₂}{\ensuremath{_2}}
\newunicodechar{ᵣ}{\ensuremath{_r}}
\newunicodechar{ₛ}{\ensuremath{_s}}
\newunicodechar{ˢ}{\ensuremath{^s}}
\newunicodechar{ʷ}{\ensuremath{^w}}
\newunicodechar{ᶜ}{\ensuremath{^c}}
\newunicodechar{∘}{\ensuremath{\circ}}
\newunicodechar{𝒜}{\ensuremath{\mathscr{A}}}
\newunicodechar{ℬ}{\ensuremath{\mathscr{B}}}
\newunicodechar{𝒞}{\ensuremath{\mathscr{C}}}
\newunicodechar{𝒟}{\ensuremath{\mathscr{D}}}
\newunicodechar{𝓥}{\ensuremath{\mathcal{V}}}
\newunicodechar{𝓣}{\ensuremath{\mathscr{T}}}
\newunicodechar{𝑓}{\ensuremath{\mathit{f}}}
\newunicodechar{𝑔}{\ensuremath{\mathit{g}}}
\newunicodechar{ℭ}{\ensuremath{\mathfrak{C}}}
\newunicodechar{𝔇}{\ensuremath{\mathfrak{D}}}
\newunicodechar{⟦}{\ensuremath{[\![}}
\newunicodechar{⟧}{\ensuremath{]\!]}}
\newunicodechar{⦃}{\ensuremath{\{\!|}}
\newunicodechar{⦄}{\ensuremath{|\!\}}}
\newunicodechar{⇛}{\ensuremath{\Longrightarrow}}
\newunicodechar{ƛ}{\ensuremath{\lambda}}
\newunicodechar{∙}{\ensuremath{\cdot}}
\newunicodechar{≡}{\ensuremath{\equiv}}

\title{Incoherent coherences}
\author{Trebor}
\date{}

\begin{document}
\maketitle
\begin{abstract}
This article explores a generic framework of well-typed and well-scoped syntaxes,
with a signature-axiom approach resembling traditional abstract algebra.
The boilerplate code needed in defining operations on syntaxes is identified
and abstracted away. Some of the frequent boilerplate proofs are also
generalized.
\end{abstract}

This is a literate Agda file, meaning that the code below
is all actually checked and automatically typeset by Agda.
The complete source code can be found at
\url{https://github.com/Trebor-Huang/STLC}.

\begin{code}[hide]%
\>[0]\AgdaSymbol{\{-\#}\AgdaSpace{}%
\AgdaKeyword{OPTIONS}\AgdaSpace{}%
\AgdaPragma{--allow-unsolved-metas}\AgdaSpace{}%
\AgdaSymbol{\#-\}}\<%
\\
\>[0]\AgdaKeyword{module}\AgdaSpace{}%
\AgdaModule{coh}\AgdaSpace{}%
\AgdaKeyword{where}\<%
\\
\>[0]\AgdaKeyword{open}\AgdaSpace{}%
\AgdaKeyword{import}\AgdaSpace{}%
\AgdaModule{Preliminaries}\<%
\\
\>[0]\AgdaKeyword{open}\AgdaSpace{}%
\AgdaKeyword{import}\AgdaSpace{}%
\AgdaModule{Agda.Primitive}\<%
\end{code}

\section{Conor's Exercise}
\begin{code}%
\>[0]\AgdaKeyword{module}\AgdaSpace{}%
\AgdaModule{Conor's-Exercise}\AgdaSpace{}%
\AgdaKeyword{where}\<%
\end{code}

It all starts when one tries to implement \emph{typed} $\lambda$-calculus
in a language with dependent types.

\AgdaTarget{𝕋}
\begin{code}%
\>[0][@{}l@{\AgdaIndent{1}}]%
\>[4]\AgdaKeyword{data}\AgdaSpace{}%
\AgdaDatatype{𝕋}\AgdaSpace{}%
\AgdaSymbol{:}\AgdaSpace{}%
\AgdaPrimitive{Set}\AgdaSpace{}%
\AgdaKeyword{where}\<%
\\
\>[4][@{}l@{\AgdaIndent{0}}]%
\>[8]\AgdaInductiveConstructor{ι}%
\>[13]\AgdaSymbol{:}\AgdaSpace{}%
\AgdaDatatype{𝕋}\<%
\\
\>[8]\AgdaOperator{\AgdaInductiveConstructor{\AgdaUnderscore{}⟶\AgdaUnderscore{}}}%
\>[13]\AgdaSymbol{:}\AgdaSpace{}%
\AgdaDatatype{𝕋}\AgdaSpace{}%
\AgdaSymbol{→}\AgdaSpace{}%
\AgdaDatatype{𝕋}\AgdaSpace{}%
\AgdaSymbol{→}\AgdaSpace{}%
\AgdaDatatype{𝕋}\<%
\\
\>[4]\AgdaKeyword{infixr}\AgdaSpace{}%
\AgdaNumber{15}\AgdaSpace{}%
\AgdaOperator{\AgdaInductiveConstructor{\AgdaUnderscore{}⟶\AgdaUnderscore{}}}\<%
\end{code}

Note that we use the same glyph $\to$ for the Agda function space and
the $\lambda$-calculus function space. The colors are different, but even
without the colors, it should be clear from context.

A context (no pun intended) is simply a snoc-list:

\begin{code}%
\>[4]\AgdaFunction{Context}\AgdaSpace{}%
\AgdaSymbol{=}\AgdaSpace{}%
\AgdaDatatype{List}\AgdaSpace{}%
\AgdaDatatype{𝕋}\<%
\\
\>[4]\AgdaFunction{\AgdaUnderscore{}}\AgdaSpace{}%
\AgdaSymbol{:}\AgdaSpace{}%
\AgdaFunction{Context}\<%
\\
\>[4]\AgdaSymbol{\AgdaUnderscore{}}\AgdaSpace{}%
\AgdaSymbol{=}\AgdaSpace{}%
\AgdaInductiveConstructor{∅}\AgdaSpace{}%
\AgdaOperator{\AgdaInductiveConstructor{◂}}\AgdaSpace{}%
\AgdaInductiveConstructor{ι}\AgdaSpace{}%
\AgdaOperator{\AgdaInductiveConstructor{◂}}\AgdaSpace{}%
\AgdaSymbol{(}\AgdaInductiveConstructor{ι}\AgdaSpace{}%
\AgdaOperator{\AgdaInductiveConstructor{⟶}}\AgdaSpace{}%
\AgdaInductiveConstructor{ι}\AgdaSymbol{)}\<%
\end{code}

We use \AgdaInductiveConstructor{\_◂\_} for a visual reminder that
it is a snoc-list.

\begin{code}[hide]%
\>[4]\AgdaKeyword{variable}\<%
\\
\>[4][@{}l@{\AgdaIndent{0}}]%
\>[8]\AgdaGeneralizable{σ}\AgdaSpace{}%
\AgdaGeneralizable{τ}\AgdaSpace{}%
\AgdaSymbol{:}\AgdaSpace{}%
\AgdaDatatype{𝕋}\<%
\\
\>[8]\AgdaGeneralizable{Γ}\AgdaSpace{}%
\AgdaGeneralizable{Δ}\AgdaSpace{}%
\AgdaGeneralizable{Ξ}\AgdaSpace{}%
\AgdaGeneralizable{Θ}\AgdaSpace{}%
\AgdaSymbol{:}\AgdaSpace{}%
\AgdaFunction{Context}\<%
\end{code}

In this way, we can ensure that our variables are well-typed:

\begin{code}%
\>[4]\AgdaKeyword{infix}\AgdaSpace{}%
\AgdaNumber{5}\AgdaSpace{}%
\AgdaOperator{\AgdaDatatype{\AgdaUnderscore{}∋\AgdaUnderscore{}}}\<%
\\
\>[4]\AgdaKeyword{data}\AgdaSpace{}%
\AgdaOperator{\AgdaDatatype{\AgdaUnderscore{}∋\AgdaUnderscore{}}}\AgdaSpace{}%
\AgdaSymbol{:}\AgdaSpace{}%
\AgdaFunction{Context}\AgdaSpace{}%
\AgdaSymbol{→}\AgdaSpace{}%
\AgdaDatatype{𝕋}\AgdaSpace{}%
\AgdaSymbol{→}\AgdaSpace{}%
\AgdaPrimitive{Set}\AgdaSpace{}%
\AgdaKeyword{where}\<%
\\
\>[4][@{}l@{\AgdaIndent{0}}]%
\>[8]\AgdaInductiveConstructor{𝕫}%
\>[11]\AgdaSymbol{:}%
\>[24]\AgdaGeneralizable{Γ}\AgdaSpace{}%
\AgdaOperator{\AgdaInductiveConstructor{◂}}\AgdaSpace{}%
\AgdaGeneralizable{σ}\AgdaSpace{}%
\AgdaOperator{\AgdaDatatype{∋}}\AgdaSpace{}%
\AgdaGeneralizable{σ}\<%
\\
\>[8]\AgdaOperator{\AgdaInductiveConstructor{𝕤\AgdaUnderscore{}}}\AgdaSpace{}%
\AgdaSymbol{:}%
\>[15]\AgdaGeneralizable{Γ}\AgdaSpace{}%
\AgdaOperator{\AgdaDatatype{∋}}\AgdaSpace{}%
\AgdaGeneralizable{τ}\AgdaSpace{}%
\AgdaSymbol{→}%
\>[24]\AgdaGeneralizable{Γ}\AgdaSpace{}%
\AgdaOperator{\AgdaInductiveConstructor{◂}}\AgdaSpace{}%
\AgdaGeneralizable{σ}\AgdaSpace{}%
\AgdaOperator{\AgdaDatatype{∋}}\AgdaSpace{}%
\AgdaGeneralizable{τ}\<%
\end{code}

Now we can define well-typed terms:

\begin{code}%
\>[4]\AgdaKeyword{infix}\AgdaSpace{}%
\AgdaNumber{5}\AgdaSpace{}%
\AgdaOperator{\AgdaDatatype{\AgdaUnderscore{}⊢\AgdaUnderscore{}}}\<%
\\
\>[4]\AgdaKeyword{data}\AgdaSpace{}%
\AgdaOperator{\AgdaDatatype{\AgdaUnderscore{}⊢\AgdaUnderscore{}}}\AgdaSpace{}%
\AgdaSymbol{:}\AgdaSpace{}%
\AgdaFunction{Context}\AgdaSpace{}%
\AgdaSymbol{→}\AgdaSpace{}%
\AgdaDatatype{𝕋}\AgdaSpace{}%
\AgdaSymbol{→}\AgdaSpace{}%
\AgdaPrimitive{Set}\AgdaSpace{}%
\AgdaKeyword{where}\<%
\\
\>[4][@{}l@{\AgdaIndent{0}}]%
\>[8]\AgdaInductiveConstructor{var}%
\>[13]\AgdaSymbol{:}%
\>[20]\AgdaGeneralizable{Γ}%
\>[23]\AgdaOperator{\AgdaDatatype{∋}}\AgdaSpace{}%
\AgdaGeneralizable{σ}%
\>[44]\AgdaSymbol{→}\AgdaSpace{}%
\AgdaGeneralizable{Γ}\AgdaSpace{}%
\AgdaOperator{\AgdaDatatype{⊢}}\AgdaSpace{}%
\AgdaGeneralizable{σ}\<%
\\
\>[8]\AgdaInductiveConstructor{app}%
\>[13]\AgdaSymbol{:}%
\>[20]\AgdaGeneralizable{Γ}%
\>[23]\AgdaOperator{\AgdaDatatype{⊢}}\AgdaSpace{}%
\AgdaGeneralizable{σ}\AgdaSpace{}%
\AgdaOperator{\AgdaInductiveConstructor{⟶}}\AgdaSpace{}%
\AgdaGeneralizable{τ}%
\>[35]\AgdaSymbol{→}\AgdaSpace{}%
\AgdaGeneralizable{Γ}\AgdaSpace{}%
\AgdaOperator{\AgdaDatatype{⊢}}\AgdaSpace{}%
\AgdaGeneralizable{σ}%
\>[44]\AgdaSymbol{→}\AgdaSpace{}%
\AgdaGeneralizable{Γ}\AgdaSpace{}%
\AgdaOperator{\AgdaDatatype{⊢}}\AgdaSpace{}%
\AgdaGeneralizable{τ}\<%
\\
\>[8]\AgdaInductiveConstructor{lam}%
\>[13]\AgdaSymbol{:}\AgdaSpace{}%
\AgdaGeneralizable{Γ}\AgdaSpace{}%
\AgdaOperator{\AgdaInductiveConstructor{◂}}%
\>[20]\AgdaGeneralizable{σ}%
\>[23]\AgdaOperator{\AgdaDatatype{⊢}}\AgdaSpace{}%
\AgdaGeneralizable{τ}%
\>[44]\AgdaSymbol{→}\AgdaSpace{}%
\AgdaGeneralizable{Γ}\AgdaSpace{}%
\AgdaOperator{\AgdaDatatype{⊢}}\AgdaSpace{}%
\AgdaGeneralizable{σ}\AgdaSpace{}%
\AgdaOperator{\AgdaInductiveConstructor{⟶}}\AgdaSpace{}%
\AgdaGeneralizable{τ}\<%
\end{code}

The type system of Agda now ensures that every term is well-typed,
blurring the distinction between syntax --- what terms are well-formed,
and semantics --- what meanings the terms have. Let's look at some
examples:

\begin{code}%
\>[4]\AgdaFunction{I}\AgdaSpace{}%
\AgdaSymbol{:}\AgdaSpace{}%
\AgdaInductiveConstructor{∅}\AgdaSpace{}%
\AgdaOperator{\AgdaDatatype{⊢}}\AgdaSpace{}%
\AgdaGeneralizable{σ}\AgdaSpace{}%
\AgdaOperator{\AgdaInductiveConstructor{⟶}}\AgdaSpace{}%
\AgdaGeneralizable{σ}\<%
\\
\>[4]\AgdaFunction{I}\AgdaSpace{}%
\AgdaSymbol{=}\AgdaSpace{}%
\AgdaInductiveConstructor{lam}\AgdaSpace{}%
\AgdaSymbol{(}\AgdaInductiveConstructor{var}\AgdaSpace{}%
\AgdaInductiveConstructor{𝕫}\AgdaSymbol{)}\<%
\\
\\[\AgdaEmptyExtraSkip]%
\>[4]\AgdaFunction{K}\AgdaSpace{}%
\AgdaSymbol{:}\AgdaSpace{}%
\AgdaInductiveConstructor{∅}\AgdaSpace{}%
\AgdaOperator{\AgdaDatatype{⊢}}\AgdaSpace{}%
\AgdaGeneralizable{σ}\AgdaSpace{}%
\AgdaOperator{\AgdaInductiveConstructor{⟶}}\AgdaSpace{}%
\AgdaGeneralizable{τ}\AgdaSpace{}%
\AgdaOperator{\AgdaInductiveConstructor{⟶}}\AgdaSpace{}%
\AgdaGeneralizable{σ}\<%
\\
\>[4]\AgdaFunction{K}\AgdaSpace{}%
\AgdaSymbol{=}\AgdaSpace{}%
\AgdaInductiveConstructor{lam}\AgdaSpace{}%
\AgdaSymbol{(}\AgdaInductiveConstructor{lam}\AgdaSpace{}%
\AgdaSymbol{(}\AgdaInductiveConstructor{var}\AgdaSpace{}%
\AgdaSymbol{(}\AgdaOperator{\AgdaInductiveConstructor{𝕤}}\AgdaSpace{}%
\AgdaInductiveConstructor{𝕫}\AgdaSymbol{)))}\<%
\end{code}

So far so good. But just \emph{defining} terms is surely not enough,
we need to be able to \emph{manipulate} them. The most fundamental
operation on syntax with variable bindings, is substitution:

\begin{code}%
\>[4]\AgdaFunction{Substitution}\AgdaSpace{}%
\AgdaSymbol{:}\AgdaSpace{}%
\AgdaFunction{Context}\AgdaSpace{}%
\AgdaSymbol{→}\AgdaSpace{}%
\AgdaFunction{Context}\AgdaSpace{}%
\AgdaSymbol{→}\AgdaSpace{}%
\AgdaPrimitive{Set}\<%
\\
\>[4]\AgdaFunction{Substitution}\AgdaSpace{}%
\AgdaBound{Γ}\AgdaSpace{}%
\AgdaBound{Δ}\AgdaSpace{}%
\AgdaSymbol{=}\AgdaSpace{}%
\AgdaSymbol{∀}\AgdaSpace{}%
\AgdaSymbol{\{}\AgdaBound{σ}\AgdaSymbol{\}}\AgdaSpace{}%
\AgdaSymbol{→}\AgdaSpace{}%
\AgdaBound{Γ}\AgdaSpace{}%
\AgdaOperator{\AgdaDatatype{∋}}\AgdaSpace{}%
\AgdaBound{σ}\AgdaSpace{}%
\AgdaSymbol{→}\AgdaSpace{}%
\AgdaBound{Δ}\AgdaSpace{}%
\AgdaOperator{\AgdaDatatype{⊢}}\AgdaSpace{}%
\AgdaBound{σ}\<%
\\
\\[\AgdaEmptyExtraSkip]%
\>[4]\AgdaFunction{Transformation}\AgdaSpace{}%
\AgdaSymbol{:}\AgdaSpace{}%
\AgdaFunction{Context}\AgdaSpace{}%
\AgdaSymbol{→}\AgdaSpace{}%
\AgdaFunction{Context}\AgdaSpace{}%
\AgdaSymbol{→}\AgdaSpace{}%
\AgdaPrimitive{Set}\<%
\\
\>[4]\AgdaFunction{Transformation}\AgdaSpace{}%
\AgdaBound{Γ}\AgdaSpace{}%
\AgdaBound{Δ}\AgdaSpace{}%
\AgdaSymbol{=}\AgdaSpace{}%
\AgdaSymbol{∀}\AgdaSpace{}%
\AgdaSymbol{\{}\AgdaBound{σ}\AgdaSymbol{\}}\AgdaSpace{}%
\AgdaSymbol{→}\AgdaSpace{}%
\AgdaBound{Γ}\AgdaSpace{}%
\AgdaOperator{\AgdaDatatype{⊢}}\AgdaSpace{}%
\AgdaBound{σ}\AgdaSpace{}%
\AgdaSymbol{→}\AgdaSpace{}%
\AgdaBound{Δ}\AgdaSpace{}%
\AgdaOperator{\AgdaDatatype{⊢}}\AgdaSpace{}%
\AgdaBound{σ}\<%
\end{code}

Let's go right into it:

\begin{code}%
\>[4]\AgdaFunction{subs₁}\AgdaSpace{}%
\AgdaSymbol{:}\AgdaSpace{}%
\AgdaFunction{Substitution}\AgdaSpace{}%
\AgdaGeneralizable{Γ}\AgdaSpace{}%
\AgdaGeneralizable{Δ}\AgdaSpace{}%
\AgdaSymbol{→}\AgdaSpace{}%
\AgdaFunction{Transformation}\AgdaSpace{}%
\AgdaGeneralizable{Γ}\AgdaSpace{}%
\AgdaGeneralizable{Δ}\<%
\\
\>[4]\AgdaComment{----\ subscript₁\ for\ "first\ attempt"}\<%
\\
\>[4]\AgdaFunction{subs₁}\AgdaSpace{}%
\AgdaBound{s}\AgdaSpace{}%
\AgdaSymbol{(}\AgdaInductiveConstructor{var}\AgdaSpace{}%
\AgdaBound{i}\AgdaSymbol{)}%
\>[25]\AgdaSymbol{=}\AgdaSpace{}%
\AgdaBound{s}\AgdaSpace{}%
\AgdaBound{i}\<%
\\
\>[4]\AgdaFunction{subs₁}\AgdaSpace{}%
\AgdaBound{s}\AgdaSpace{}%
\AgdaSymbol{(}\AgdaInductiveConstructor{app}\AgdaSpace{}%
\AgdaBound{t₁}\AgdaSpace{}%
\AgdaBound{t₂}\AgdaSymbol{)}%
\>[25]\AgdaSymbol{=}\AgdaSpace{}%
\AgdaInductiveConstructor{app}\AgdaSpace{}%
\AgdaSymbol{(}\AgdaFunction{subs₁}\AgdaSpace{}%
\AgdaBound{s}\AgdaSpace{}%
\AgdaBound{t₁}\AgdaSymbol{)}\AgdaSpace{}%
\AgdaSymbol{(}\AgdaFunction{subs₁}\AgdaSpace{}%
\AgdaBound{s}\AgdaSpace{}%
\AgdaBound{t₂}\AgdaSymbol{)}\<%
\end{code}

The powerful type system helps us write the program: By asking
Agda to do a case-split on the term being transformed, we immediately
get the required branches. Agda can also automatically generate the
correct term to write in the first two branches. But in the third branch
there is a problem:

\begin{code}%
\>[4]\AgdaFunction{subs₁}\AgdaSpace{}%
\AgdaBound{s}\AgdaSpace{}%
\AgdaSymbol{(}\AgdaInductiveConstructor{lam}\AgdaSpace{}%
\AgdaBound{t}\AgdaSymbol{)}\AgdaSpace{}%
\AgdaSymbol{=}\AgdaSpace{}%
\AgdaInductiveConstructor{lam}\AgdaSpace{}%
\AgdaSymbol{(}\AgdaFunction{subs₁}\AgdaSpace{}%
\AgdaHole{\{!\ \ \ !\}}\AgdaSpace{}%
\AgdaBound{t}\AgdaSymbol{)}\<%
\end{code}

We need a way to ``push in'' an extra variable. And here we go:

\begin{code}%
\>[4]\AgdaFunction{push₁}\AgdaSpace{}%
\AgdaSymbol{:}\AgdaSpace{}%
\AgdaFunction{Substitution}\AgdaSpace{}%
\AgdaGeneralizable{Γ}\AgdaSpace{}%
\AgdaGeneralizable{Δ}\AgdaSpace{}%
\AgdaSymbol{→}\AgdaSpace{}%
\AgdaFunction{Substitution}\AgdaSpace{}%
\AgdaSymbol{(}\AgdaGeneralizable{Γ}\AgdaSpace{}%
\AgdaOperator{\AgdaInductiveConstructor{◂}}\AgdaSpace{}%
\AgdaGeneralizable{σ}\AgdaSymbol{)}\AgdaSpace{}%
\AgdaSymbol{(}\AgdaGeneralizable{Δ}\AgdaSpace{}%
\AgdaOperator{\AgdaInductiveConstructor{◂}}\AgdaSpace{}%
\AgdaGeneralizable{σ}\AgdaSymbol{)}\<%
\\
\>[4]\AgdaFunction{push₁}\AgdaSpace{}%
\AgdaBound{σ}\AgdaSpace{}%
\AgdaInductiveConstructor{𝕫}%
\>[19]\AgdaSymbol{=}\AgdaSpace{}%
\AgdaInductiveConstructor{var}\AgdaSpace{}%
\AgdaInductiveConstructor{𝕫}\<%
\\
\>[4]\AgdaFunction{push₁}\AgdaSpace{}%
\AgdaBound{σ}\AgdaSpace{}%
\AgdaSymbol{(}\AgdaOperator{\AgdaInductiveConstructor{𝕤}}\AgdaSpace{}%
\AgdaBound{i}\AgdaSymbol{)}%
\>[19]\AgdaSymbol{=}\AgdaSpace{}%
\AgdaHole{\{!\ \ \ !\}}\<%
\end{code}

This in turn requires us to weaken a term:

\begin{code}%
\>[4]\AgdaFunction{weaken₁}\AgdaSpace{}%
\AgdaSymbol{:}\AgdaSpace{}%
\AgdaGeneralizable{Γ}\AgdaSpace{}%
\AgdaOperator{\AgdaDatatype{⊢}}\AgdaSpace{}%
\AgdaGeneralizable{σ}\AgdaSpace{}%
\AgdaSymbol{→}\AgdaSpace{}%
\AgdaGeneralizable{Γ}\AgdaSpace{}%
\AgdaOperator{\AgdaInductiveConstructor{◂}}\AgdaSpace{}%
\AgdaGeneralizable{τ}\AgdaSpace{}%
\AgdaOperator{\AgdaDatatype{⊢}}\AgdaSpace{}%
\AgdaGeneralizable{σ}\<%
\\
\>[4]\AgdaFunction{weaken₁}\AgdaSpace{}%
\AgdaSymbol{(}\AgdaInductiveConstructor{var}\AgdaSpace{}%
\AgdaBound{i}\AgdaSymbol{)}%
\>[25]\AgdaSymbol{=}\AgdaSpace{}%
\AgdaInductiveConstructor{var}\AgdaSpace{}%
\AgdaSymbol{(}\AgdaOperator{\AgdaInductiveConstructor{𝕤}}\AgdaSpace{}%
\AgdaBound{i}\AgdaSymbol{)}\<%
\\
\>[4]\AgdaFunction{weaken₁}\AgdaSpace{}%
\AgdaSymbol{(}\AgdaInductiveConstructor{app}\AgdaSpace{}%
\AgdaBound{t₁}\AgdaSpace{}%
\AgdaBound{t₂}\AgdaSymbol{)}%
\>[25]\AgdaSymbol{=}\AgdaSpace{}%
\AgdaInductiveConstructor{app}\AgdaSpace{}%
\AgdaSymbol{(}\AgdaFunction{weaken₁}\AgdaSpace{}%
\AgdaBound{t₁}\AgdaSymbol{)}\AgdaSpace{}%
\AgdaSymbol{(}\AgdaFunction{weaken₁}\AgdaSpace{}%
\AgdaBound{t₂}\AgdaSymbol{)}\<%
\\
\>[4]\AgdaFunction{weaken₁}\AgdaSpace{}%
\AgdaSymbol{(}\AgdaInductiveConstructor{lam}\AgdaSpace{}%
\AgdaBound{t}\AgdaSymbol{)}%
\>[25]\AgdaSymbol{=}\AgdaSpace{}%
\AgdaInductiveConstructor{lam}\AgdaSpace{}%
\AgdaHole{\{!\ \ \ !\}}\<%
\end{code}

The first two cases are still easy, but the \AgdaInductiveConstructor{lam}
case is problematic. We need to push in yet another variable!

It turns out that we need to do this in two steps. First, we deal with
variable renamings only:

\begin{code}%
\>[4]\AgdaFunction{Renaming}\AgdaSpace{}%
\AgdaSymbol{:}\AgdaSpace{}%
\AgdaFunction{Context}\AgdaSpace{}%
\AgdaSymbol{→}\AgdaSpace{}%
\AgdaFunction{Context}\AgdaSpace{}%
\AgdaSymbol{→}\AgdaSpace{}%
\AgdaPrimitive{Set}\<%
\\
\>[4]\AgdaFunction{Renaming}\AgdaSpace{}%
\AgdaBound{Γ}\AgdaSpace{}%
\AgdaBound{Δ}\AgdaSpace{}%
\AgdaSymbol{=}\AgdaSpace{}%
\AgdaSymbol{∀}\AgdaSpace{}%
\AgdaSymbol{\{}\AgdaBound{σ}\AgdaSymbol{\}}\AgdaSpace{}%
\AgdaSymbol{→}\AgdaSpace{}%
\AgdaBound{Γ}\AgdaSpace{}%
\AgdaOperator{\AgdaDatatype{∋}}\AgdaSpace{}%
\AgdaBound{σ}\AgdaSpace{}%
\AgdaSymbol{→}\AgdaSpace{}%
\AgdaBound{Δ}\AgdaSpace{}%
\AgdaOperator{\AgdaDatatype{∋}}\AgdaSpace{}%
\AgdaBound{σ}\<%
\\
\\[\AgdaEmptyExtraSkip]%
\>[4]\AgdaFunction{weakenᵣ}\AgdaSpace{}%
\AgdaSymbol{:}\AgdaSpace{}%
\AgdaFunction{Renaming}\AgdaSpace{}%
\AgdaGeneralizable{Γ}\AgdaSpace{}%
\AgdaGeneralizable{Δ}\AgdaSpace{}%
\AgdaSymbol{→}\AgdaSpace{}%
\AgdaFunction{Renaming}\AgdaSpace{}%
\AgdaSymbol{(}\AgdaGeneralizable{Γ}\AgdaSpace{}%
\AgdaOperator{\AgdaInductiveConstructor{◂}}\AgdaSpace{}%
\AgdaGeneralizable{σ}\AgdaSymbol{)}\AgdaSpace{}%
\AgdaSymbol{(}\AgdaGeneralizable{Δ}\AgdaSpace{}%
\AgdaOperator{\AgdaInductiveConstructor{◂}}\AgdaSpace{}%
\AgdaGeneralizable{σ}\AgdaSymbol{)}\<%
\\
\>[4]\AgdaFunction{weakenᵣ}\AgdaSpace{}%
\AgdaBound{r}\AgdaSpace{}%
\AgdaInductiveConstructor{𝕫}%
\>[21]\AgdaSymbol{=}\AgdaSpace{}%
\AgdaInductiveConstructor{𝕫}\<%
\\
\>[4]\AgdaFunction{weakenᵣ}\AgdaSpace{}%
\AgdaBound{r}\AgdaSpace{}%
\AgdaSymbol{(}\AgdaOperator{\AgdaInductiveConstructor{𝕤}}\AgdaSpace{}%
\AgdaBound{i}\AgdaSymbol{)}%
\>[21]\AgdaSymbol{=}\AgdaSpace{}%
\AgdaOperator{\AgdaInductiveConstructor{𝕤}}\AgdaSpace{}%
\AgdaBound{r}\AgdaSpace{}%
\AgdaBound{i}\<%
\\
\\[\AgdaEmptyExtraSkip]%
\>[4]\AgdaFunction{rename}\AgdaSpace{}%
\AgdaSymbol{:}\AgdaSpace{}%
\AgdaFunction{Renaming}\AgdaSpace{}%
\AgdaGeneralizable{Γ}\AgdaSpace{}%
\AgdaGeneralizable{Δ}\AgdaSpace{}%
\AgdaSymbol{→}\AgdaSpace{}%
\AgdaFunction{Transformation}\AgdaSpace{}%
\AgdaGeneralizable{Γ}\AgdaSpace{}%
\AgdaGeneralizable{Δ}\<%
\\
\>[4]\AgdaFunction{rename}\AgdaSpace{}%
\AgdaBound{r}\AgdaSpace{}%
\AgdaSymbol{(}\AgdaInductiveConstructor{var}\AgdaSpace{}%
\AgdaBound{i}\AgdaSymbol{)}%
\>[26]\AgdaSymbol{=}\AgdaSpace{}%
\AgdaInductiveConstructor{var}\AgdaSpace{}%
\AgdaSymbol{(}\AgdaBound{r}\AgdaSpace{}%
\AgdaBound{i}\AgdaSymbol{)}\<%
\\
\>[4]\AgdaFunction{rename}\AgdaSpace{}%
\AgdaBound{r}\AgdaSpace{}%
\AgdaSymbol{(}\AgdaInductiveConstructor{app}\AgdaSpace{}%
\AgdaBound{t₁}\AgdaSpace{}%
\AgdaBound{t₂}\AgdaSymbol{)}%
\>[26]\AgdaSymbol{=}\AgdaSpace{}%
\AgdaInductiveConstructor{app}\AgdaSpace{}%
\AgdaSymbol{(}\AgdaFunction{rename}\AgdaSpace{}%
\AgdaBound{r}\AgdaSpace{}%
\AgdaBound{t₁}\AgdaSymbol{)}\AgdaSpace{}%
\AgdaSymbol{(}\AgdaFunction{rename}\AgdaSpace{}%
\AgdaBound{r}\AgdaSpace{}%
\AgdaBound{t₂}\AgdaSymbol{)}\<%
\\
\>[4]\AgdaFunction{rename}\AgdaSpace{}%
\AgdaBound{r}\AgdaSpace{}%
\AgdaSymbol{(}\AgdaInductiveConstructor{lam}\AgdaSpace{}%
\AgdaBound{t}\AgdaSymbol{)}%
\>[26]\AgdaSymbol{=}\AgdaSpace{}%
\AgdaInductiveConstructor{lam}\AgdaSpace{}%
\AgdaSymbol{(}\AgdaFunction{rename}\AgdaSpace{}%
\AgdaSymbol{(}\AgdaFunction{weakenᵣ}\AgdaSpace{}%
\AgdaBound{r}\AgdaSymbol{)}\AgdaSpace{}%
\AgdaBound{t}\AgdaSymbol{)}\<%
\end{code}

The \AgdaInductiveConstructor{lam} case now goes through. And we can
finish off the development:

\begin{code}%
\>[4]\AgdaFunction{weakenₛ}\AgdaSpace{}%
\AgdaSymbol{:}\AgdaSpace{}%
\AgdaFunction{Substitution}\AgdaSpace{}%
\AgdaGeneralizable{Γ}\AgdaSpace{}%
\AgdaGeneralizable{Δ}\AgdaSpace{}%
\AgdaSymbol{→}\AgdaSpace{}%
\AgdaFunction{Substitution}\AgdaSpace{}%
\AgdaSymbol{(}\AgdaGeneralizable{Γ}\AgdaSpace{}%
\AgdaOperator{\AgdaInductiveConstructor{◂}}\AgdaSpace{}%
\AgdaGeneralizable{σ}\AgdaSymbol{)}\AgdaSpace{}%
\AgdaSymbol{(}\AgdaGeneralizable{Δ}\AgdaSpace{}%
\AgdaOperator{\AgdaInductiveConstructor{◂}}\AgdaSpace{}%
\AgdaGeneralizable{σ}\AgdaSymbol{)}\<%
\\
\>[4]\AgdaFunction{weakenₛ}\AgdaSpace{}%
\AgdaBound{s}\AgdaSpace{}%
\AgdaInductiveConstructor{𝕫}%
\>[21]\AgdaSymbol{=}\AgdaSpace{}%
\AgdaInductiveConstructor{var}\AgdaSpace{}%
\AgdaInductiveConstructor{𝕫}\<%
\\
\>[4]\AgdaFunction{weakenₛ}\AgdaSpace{}%
\AgdaBound{s}\AgdaSpace{}%
\AgdaSymbol{(}\AgdaOperator{\AgdaInductiveConstructor{𝕤}}\AgdaSpace{}%
\AgdaBound{i}\AgdaSymbol{)}%
\>[21]\AgdaSymbol{=}\AgdaSpace{}%
\AgdaFunction{rename}\AgdaSpace{}%
\AgdaOperator{\AgdaInductiveConstructor{𝕤\AgdaUnderscore{}}}\AgdaSpace{}%
\AgdaSymbol{(}\AgdaBound{s}\AgdaSpace{}%
\AgdaBound{i}\AgdaSymbol{)}\<%
\\
\\[\AgdaEmptyExtraSkip]%
\>[4]\AgdaFunction{subs}\AgdaSpace{}%
\AgdaSymbol{:}\AgdaSpace{}%
\AgdaFunction{Substitution}\AgdaSpace{}%
\AgdaGeneralizable{Γ}\AgdaSpace{}%
\AgdaGeneralizable{Δ}\AgdaSpace{}%
\AgdaSymbol{→}\AgdaSpace{}%
\AgdaFunction{Transformation}\AgdaSpace{}%
\AgdaGeneralizable{Γ}\AgdaSpace{}%
\AgdaGeneralizable{Δ}\<%
\\
\>[4]\AgdaFunction{subs}\AgdaSpace{}%
\AgdaBound{s}\AgdaSpace{}%
\AgdaSymbol{(}\AgdaInductiveConstructor{var}\AgdaSpace{}%
\AgdaBound{i}\AgdaSymbol{)}%
\>[24]\AgdaSymbol{=}\AgdaSpace{}%
\AgdaBound{s}\AgdaSpace{}%
\AgdaBound{i}\<%
\\
\>[4]\AgdaFunction{subs}\AgdaSpace{}%
\AgdaBound{s}\AgdaSpace{}%
\AgdaSymbol{(}\AgdaInductiveConstructor{app}\AgdaSpace{}%
\AgdaBound{t₁}\AgdaSpace{}%
\AgdaBound{t₂}\AgdaSymbol{)}%
\>[24]\AgdaSymbol{=}\AgdaSpace{}%
\AgdaInductiveConstructor{app}\AgdaSpace{}%
\AgdaSymbol{(}\AgdaFunction{subs}\AgdaSpace{}%
\AgdaBound{s}\AgdaSpace{}%
\AgdaBound{t₁}\AgdaSymbol{)}\AgdaSpace{}%
\AgdaSymbol{(}\AgdaFunction{subs}\AgdaSpace{}%
\AgdaBound{s}\AgdaSpace{}%
\AgdaBound{t₂}\AgdaSymbol{)}\<%
\\
\>[4]\AgdaFunction{subs}\AgdaSpace{}%
\AgdaBound{s}\AgdaSpace{}%
\AgdaSymbol{(}\AgdaInductiveConstructor{lam}\AgdaSpace{}%
\AgdaBound{t}\AgdaSymbol{)}%
\>[24]\AgdaSymbol{=}\AgdaSpace{}%
\AgdaInductiveConstructor{lam}\AgdaSpace{}%
\AgdaSymbol{(}\AgdaFunction{subs}\AgdaSpace{}%
\AgdaSymbol{(}\AgdaFunction{weakenₛ}\AgdaSpace{}%
\AgdaBound{s}\AgdaSymbol{)}\AgdaSpace{}%
\AgdaBound{t}\AgdaSymbol{)}\<%
\end{code}

In retrospect, the reason that we have to do this in two steps, is that
\AgdaDatatype{\_⊢\_} is \emph{defined} in two steps: It requires \AgdaDatatype{\_∋\_}
in its definition.

However, comparing the two pairs of functions we can see some sort of pattern.
It is called \emph{weakening-then-traversal} in exercise 19 of \cite{conor}. And let's do that.

\section{Abstraction and Generality}
\begin{code}%
\>[0]\AgdaKeyword{module}\AgdaSpace{}%
\AgdaModule{Abstraction}\AgdaSpace{}%
\AgdaSymbol{(}\AgdaBound{I}\AgdaSpace{}%
\AgdaSymbol{:}\AgdaSpace{}%
\AgdaPrimitive{Set}\AgdaSymbol{)}\AgdaSpace{}%
\AgdaKeyword{where}\<%
\end{code}

In this section, we will work with an abstract parameter \AgdaArgument{I}
instead of \AgdaDatatype{𝕋} in the previous section. We can start by
noticing the similarity in the type signature:

\AgdaTarget{Scope, Morph}
\begin{code}%
\>[0][@{}l@{\AgdaIndent{1}}]%
\>[4]\AgdaFunction{Scope}\AgdaSpace{}%
\AgdaSymbol{=}\AgdaSpace{}%
\AgdaSymbol{(}\AgdaBound{Γ}\AgdaSpace{}%
\AgdaSymbol{:}\AgdaSpace{}%
\AgdaDatatype{List}\AgdaSpace{}%
\AgdaBound{I}\AgdaSymbol{)}\AgdaSpace{}%
\AgdaSymbol{(}\AgdaBound{i}\AgdaSpace{}%
\AgdaSymbol{:}\AgdaSpace{}%
\AgdaBound{I}\AgdaSymbol{)}\AgdaSpace{}%
\AgdaSymbol{→}\AgdaSpace{}%
\AgdaPrimitive{Set}\<%
\\
\>[4]\AgdaFunction{Morph}\AgdaSpace{}%
\AgdaSymbol{=}\AgdaSpace{}%
\AgdaSymbol{(}\AgdaBound{Γ}\AgdaSpace{}%
\AgdaBound{Δ}\AgdaSpace{}%
\AgdaSymbol{:}\AgdaSpace{}%
\AgdaDatatype{List}\AgdaSpace{}%
\AgdaBound{I}\AgdaSymbol{)}\AgdaSpace{}%
\AgdaSymbol{→}\AgdaSpace{}%
\AgdaPrimitive{Set}\<%
\end{code}
\begin{code}[hide]%
\>[4]\AgdaKeyword{private}\AgdaSpace{}%
\AgdaKeyword{variable}\<%
\\
\>[4][@{}l@{\AgdaIndent{0}}]%
\>[8]\AgdaGeneralizable{Γ}\AgdaSpace{}%
\AgdaGeneralizable{Δ}\AgdaSpace{}%
\AgdaGeneralizable{Θ}\AgdaSpace{}%
\AgdaGeneralizable{Ξ}\AgdaSpace{}%
\AgdaSymbol{:}\AgdaSpace{}%
\AgdaDatatype{List}\AgdaSpace{}%
\AgdaBound{I}\<%
\\
\>[8]\AgdaGeneralizable{i}\AgdaSpace{}%
\AgdaGeneralizable{j}\AgdaSpace{}%
\AgdaGeneralizable{k}\AgdaSpace{}%
\AgdaGeneralizable{l}\AgdaSpace{}%
\AgdaSymbol{:}\AgdaSpace{}%
\AgdaBound{I}\<%
\\
\>[8]\AgdaGeneralizable{𝒜}\AgdaSpace{}%
\AgdaGeneralizable{ℬ}\AgdaSpace{}%
\AgdaGeneralizable{𝒞}\AgdaSpace{}%
\AgdaGeneralizable{𝒟}\AgdaSpace{}%
\AgdaGeneralizable{𝒱}\AgdaSpace{}%
\AgdaGeneralizable{𝒲}\AgdaSpace{}%
\AgdaSymbol{:}\AgdaSpace{}%
\AgdaFunction{Scope}\<%
\end{code}

\AgdaDatatype{\_⊢\_} and \AgdaDatatype{\_∋\_} both have type \AgdaFunction{Scope}.
And \AgdaFunction{Renaming}, \AgdaFunction{Substitution} and
\AgdaFunction{Transformation} all have type \AgdaFunction{Morph}. The name
``scope'' comes from \cite{gallais}.

In the untyped case, \AgdaArgument{I} is simply the singleton type (``Untyped is
uni-typed''). This, in abstract nonsense, makes \AgdaFunction{Scope} the type of
\textbf{presheafs} on the category of renamings. \AgdaFunction{Morph} is
then natural transformations between the presheafs.%
\footnote{Of course, we haven't imposed the functorial laws yet, so
they are better described as \emph{raw} presheafs and natural transformations.
I will not spell out the categorical details, since I'm not going to
use the category-theoretic language in an essential way. More can be read at
\cite{presheaf}.}

Now we define the standard well-typed variables, which can be regarded
as the image of the Yoneda embedding:

\AgdaTarget{𝓥}
\begin{code}%
\>[4]\AgdaKeyword{infix}\AgdaSpace{}%
\AgdaNumber{5}\AgdaSpace{}%
\AgdaOperator{\AgdaDatatype{\AgdaUnderscore{}∋\AgdaUnderscore{}}}\<%
\\
\>[4]\AgdaKeyword{data}\AgdaSpace{}%
\AgdaOperator{\AgdaDatatype{\AgdaUnderscore{}∋\AgdaUnderscore{}}}\AgdaSpace{}%
\AgdaSymbol{:}\AgdaSpace{}%
\AgdaFunction{Scope}\AgdaSpace{}%
\AgdaKeyword{where}\<%
\\
\>[4][@{}l@{\AgdaIndent{0}}]%
\>[8]\AgdaInductiveConstructor{𝕫}%
\>[12]\AgdaSymbol{:}%
\>[23]\AgdaGeneralizable{Γ}\AgdaSpace{}%
\AgdaOperator{\AgdaInductiveConstructor{◂}}\AgdaSpace{}%
\AgdaGeneralizable{i}%
\>[30]\AgdaOperator{\AgdaDatatype{∋}}\AgdaSpace{}%
\AgdaGeneralizable{i}\<%
\\
\>[8]\AgdaOperator{\AgdaInductiveConstructor{𝕤\AgdaUnderscore{}}}%
\>[12]\AgdaSymbol{:}\AgdaSpace{}%
\AgdaGeneralizable{Γ}\AgdaSpace{}%
\AgdaOperator{\AgdaDatatype{∋}}\AgdaSpace{}%
\AgdaGeneralizable{i}\AgdaSpace{}%
\AgdaSymbol{→}%
\>[23]\AgdaGeneralizable{Γ}\AgdaSpace{}%
\AgdaOperator{\AgdaInductiveConstructor{◂}}\AgdaSpace{}%
\AgdaGeneralizable{j}%
\>[30]\AgdaOperator{\AgdaDatatype{∋}}\AgdaSpace{}%
\AgdaGeneralizable{i}\<%
\\
\>[4]\AgdaKeyword{infixr}\AgdaSpace{}%
\AgdaNumber{100}\AgdaSpace{}%
\AgdaOperator{\AgdaInductiveConstructor{𝕤\AgdaUnderscore{}}}\<%
\\
\\[\AgdaEmptyExtraSkip]%
\>[4]\AgdaFunction{𝓥}\AgdaSpace{}%
\AgdaSymbol{=}\AgdaSpace{}%
\AgdaOperator{\AgdaDatatype{\AgdaUnderscore{}∋\AgdaUnderscore{}}}\<%
\end{code}
\begin{code}[hide]%
\>[4]\AgdaKeyword{private}\AgdaSpace{}%
\AgdaKeyword{variable}\<%
\\
\>[4][@{}l@{\AgdaIndent{0}}]%
\>[8]\AgdaGeneralizable{v}\AgdaSpace{}%
\AgdaGeneralizable{w}\AgdaSpace{}%
\AgdaSymbol{:}\AgdaSpace{}%
\AgdaFunction{𝓥}\AgdaSpace{}%
\AgdaGeneralizable{Γ}\AgdaSpace{}%
\AgdaGeneralizable{i}\<%
\end{code}

Next, some combinators that already emerged in the last section.

\AgdaTarget{⇒, [, ], ⇛, ⟦, ⟧}
\begin{code}%
\>[4]\AgdaKeyword{infix}\AgdaSpace{}%
\AgdaNumber{4}\AgdaSpace{}%
\AgdaOperator{\AgdaFunction{\AgdaUnderscore{}⇒\AgdaUnderscore{}}}\<%
\\
\>[4]\AgdaOperator{\AgdaFunction{\AgdaUnderscore{}⇒\AgdaUnderscore{}}}\AgdaSpace{}%
\AgdaSymbol{:}\AgdaSpace{}%
\AgdaFunction{Scope}\AgdaSpace{}%
\AgdaSymbol{→}\AgdaSpace{}%
\AgdaFunction{Scope}\AgdaSpace{}%
\AgdaSymbol{→}\AgdaSpace{}%
\AgdaFunction{Morph}\<%
\\
\>[4]\AgdaSymbol{(}\AgdaBound{𝒞}\AgdaSpace{}%
\AgdaOperator{\AgdaFunction{⇒}}\AgdaSpace{}%
\AgdaBound{𝒟}\AgdaSymbol{)}\AgdaSpace{}%
\AgdaBound{Γ}\AgdaSpace{}%
\AgdaBound{Δ}\AgdaSpace{}%
\AgdaSymbol{=}\AgdaSpace{}%
\AgdaSymbol{∀}\AgdaSpace{}%
\AgdaSymbol{\{}\AgdaBound{i}\AgdaSymbol{\}}\AgdaSpace{}%
\AgdaSymbol{→}\AgdaSpace{}%
\AgdaBound{𝒞}\AgdaSpace{}%
\AgdaBound{Γ}\AgdaSpace{}%
\AgdaBound{i}\AgdaSpace{}%
\AgdaSymbol{→}\AgdaSpace{}%
\AgdaBound{𝒟}\AgdaSpace{}%
\AgdaBound{Δ}\AgdaSpace{}%
\AgdaBound{i}\<%
\\
\\[\AgdaEmptyExtraSkip]%
\>[4]\AgdaOperator{\AgdaFunction{[\AgdaUnderscore{}]}}\AgdaSpace{}%
\AgdaSymbol{:}\AgdaSpace{}%
\AgdaFunction{Morph}\AgdaSpace{}%
\AgdaSymbol{→}\AgdaSpace{}%
\AgdaPrimitive{Set}\<%
\\
\>[4]\AgdaOperator{\AgdaFunction{[}}\AgdaSpace{}%
\AgdaBound{ℭ}\AgdaSpace{}%
\AgdaOperator{\AgdaFunction{]}}\AgdaSpace{}%
\AgdaSymbol{=}\AgdaSpace{}%
\AgdaSymbol{∀}\AgdaSpace{}%
\AgdaSymbol{\{}\AgdaBound{Γ}\AgdaSymbol{\}}\AgdaSpace{}%
\AgdaSymbol{→}\AgdaSpace{}%
\AgdaBound{ℭ}\AgdaSpace{}%
\AgdaBound{Γ}\AgdaSpace{}%
\AgdaBound{Γ}\<%
\\
\\[\AgdaEmptyExtraSkip]%
\>[4]\AgdaKeyword{infixr}\AgdaSpace{}%
\AgdaNumber{3}\AgdaSpace{}%
\AgdaOperator{\AgdaFunction{\AgdaUnderscore{}⇛\AgdaUnderscore{}}}\<%
\\
\>[4]\AgdaOperator{\AgdaFunction{\AgdaUnderscore{}⇛\AgdaUnderscore{}}}\AgdaSpace{}%
\AgdaSymbol{:}\AgdaSpace{}%
\AgdaFunction{Morph}\AgdaSpace{}%
\AgdaSymbol{→}\AgdaSpace{}%
\AgdaFunction{Morph}\AgdaSpace{}%
\AgdaSymbol{→}\AgdaSpace{}%
\AgdaFunction{Morph}\<%
\\
\>[4]\AgdaSymbol{(}\AgdaBound{ℭ}\AgdaSpace{}%
\AgdaOperator{\AgdaFunction{⇛}}\AgdaSpace{}%
\AgdaBound{𝔇}\AgdaSymbol{)}\AgdaSpace{}%
\AgdaBound{Γ}\AgdaSpace{}%
\AgdaBound{Δ}\AgdaSpace{}%
\AgdaSymbol{=}\AgdaSpace{}%
\AgdaBound{ℭ}\AgdaSpace{}%
\AgdaBound{Γ}\AgdaSpace{}%
\AgdaBound{Δ}\AgdaSpace{}%
\AgdaSymbol{→}\AgdaSpace{}%
\AgdaBound{𝔇}\AgdaSpace{}%
\AgdaBound{Γ}\AgdaSpace{}%
\AgdaBound{Δ}\<%
\\
\\[\AgdaEmptyExtraSkip]%
\>[4]\AgdaOperator{\AgdaFunction{⟦\AgdaUnderscore{}⟧}}\AgdaSpace{}%
\AgdaSymbol{:}\AgdaSpace{}%
\AgdaFunction{Morph}\AgdaSpace{}%
\AgdaSymbol{→}\AgdaSpace{}%
\AgdaPrimitive{Set}\<%
\\
\>[4]\AgdaOperator{\AgdaFunction{⟦}}\AgdaSpace{}%
\AgdaBound{ℭ}\AgdaSpace{}%
\AgdaOperator{\AgdaFunction{⟧}}\AgdaSpace{}%
\AgdaSymbol{=}\AgdaSpace{}%
\AgdaSymbol{∀}\AgdaSpace{}%
\AgdaSymbol{\{}\AgdaBound{Γ}\AgdaSpace{}%
\AgdaBound{Δ}\AgdaSymbol{\}}\AgdaSpace{}%
\AgdaSymbol{→}\AgdaSpace{}%
\AgdaBound{ℭ}\AgdaSpace{}%
\AgdaBound{Γ}\AgdaSpace{}%
\AgdaBound{Δ}\<%
\end{code}

With these, we can redefine \AgdaFunction{Renaming}, \AgdaFunction{Substitution} and
\AgdaFunction{Transformation} uniformly:

\begin{code}%
\>[4]\AgdaKeyword{module}\AgdaSpace{}%
\AgdaModule{\AgdaUnderscore{}}\AgdaSpace{}%
\AgdaSymbol{(}\AgdaBound{𝒞}\AgdaSpace{}%
\AgdaSymbol{:}\AgdaSpace{}%
\AgdaFunction{Scope}\AgdaSymbol{)}\AgdaSpace{}%
\AgdaKeyword{where}\<%
\\
\>[4][@{}l@{\AgdaIndent{0}}]%
\>[8]\AgdaFunction{Renaming}\AgdaSpace{}%
\AgdaFunction{Substitution}\AgdaSpace{}%
\AgdaFunction{Transformation}\AgdaSpace{}%
\AgdaSymbol{:}\AgdaSpace{}%
\AgdaFunction{Morph}\<%
\\
\\[\AgdaEmptyExtraSkip]%
\>[8]\AgdaFunction{Renaming}%
\>[24]\AgdaSymbol{=}\AgdaSpace{}%
\AgdaFunction{𝓥}%
\>[29]\AgdaOperator{\AgdaFunction{⇒}}\AgdaSpace{}%
\AgdaFunction{𝓥}\<%
\\
\>[8]\AgdaFunction{Substitution}%
\>[24]\AgdaSymbol{=}\AgdaSpace{}%
\AgdaFunction{𝓥}%
\>[29]\AgdaOperator{\AgdaFunction{⇒}}\AgdaSpace{}%
\AgdaBound{𝒞}\<%
\\
\>[8]\AgdaFunction{Transformation}%
\>[24]\AgdaSymbol{=}\AgdaSpace{}%
\AgdaBound{𝒞}%
\>[29]\AgdaOperator{\AgdaFunction{⇒}}\AgdaSpace{}%
\AgdaBound{𝒞}\<%
\end{code}

In \cite{gallais}, a notion of ``generic syntax'' is built, and
a datatype \AgdaDatatype{Desc} is used to describe syntaxes in general.
This is very much like building a datatype \AgdaDatatype{Poly} to describe
polynomial functors, and generating the initial algebras once and for all:

\begin{code}%
\>[4]\AgdaKeyword{module}\AgdaSpace{}%
\AgdaModule{Generic}\AgdaSpace{}%
\AgdaKeyword{where}\<%
\\
\>[4][@{}l@{\AgdaIndent{0}}]%
\>[8]\AgdaKeyword{data}\AgdaSpace{}%
\AgdaDatatype{Poly}\AgdaSpace{}%
\AgdaSymbol{:}\AgdaSpace{}%
\AgdaPrimitive{Set₁}\AgdaSpace{}%
\AgdaKeyword{where}\<%
\\
\>[8][@{}l@{\AgdaIndent{0}}]%
\>[12]\AgdaInductiveConstructor{Const}%
\>[20]\AgdaSymbol{:}\AgdaSpace{}%
\AgdaPrimitive{Set}\AgdaSpace{}%
\AgdaSymbol{→}\AgdaSpace{}%
\AgdaDatatype{Poly}\<%
\\
\>[12]\AgdaInductiveConstructor{Id}%
\>[20]\AgdaSymbol{:}\AgdaSpace{}%
\AgdaDatatype{Poly}\<%
\\
\>[12]\AgdaOperator{\AgdaInductiveConstructor{\AgdaUnderscore{}⊕\AgdaUnderscore{}}}%
\>[20]\AgdaSymbol{:}\AgdaSpace{}%
\AgdaDatatype{Poly}\AgdaSpace{}%
\AgdaSymbol{→}\AgdaSpace{}%
\AgdaDatatype{Poly}\AgdaSpace{}%
\AgdaSymbol{→}\AgdaSpace{}%
\AgdaDatatype{Poly}\<%
\\
\>[12]\AgdaOperator{\AgdaInductiveConstructor{\AgdaUnderscore{}⊗\AgdaUnderscore{}}}%
\>[20]\AgdaSymbol{:}\AgdaSpace{}%
\AgdaDatatype{Poly}\AgdaSpace{}%
\AgdaSymbol{→}\AgdaSpace{}%
\AgdaDatatype{Poly}\AgdaSpace{}%
\AgdaSymbol{→}\AgdaSpace{}%
\AgdaDatatype{Poly}\<%
\\
\\[\AgdaEmptyExtraSkip]%
\>[8]\AgdaComment{----\ interprets\ Poly\ into\ Functor\ (i.e.\ Set\ →\ Set)}\<%
\\
\>[8]\AgdaFunction{interp}\AgdaSpace{}%
\AgdaSymbol{:}\AgdaSpace{}%
\AgdaDatatype{Poly}\AgdaSpace{}%
\AgdaSymbol{→}\AgdaSpace{}%
\AgdaSymbol{(}\AgdaPrimitive{Set}\AgdaSpace{}%
\AgdaSymbol{→}\AgdaSpace{}%
\AgdaPrimitive{Set}\AgdaSymbol{)}\<%
\\
\>[8]\AgdaFunction{interp}\AgdaSpace{}%
\AgdaSymbol{(}\AgdaInductiveConstructor{Const}\AgdaSpace{}%
\AgdaBound{B}\AgdaSymbol{)}\AgdaSpace{}%
\AgdaBound{A}%
\>[28]\AgdaSymbol{=}\AgdaSpace{}%
\AgdaBound{B}\<%
\\
\>[8]\AgdaFunction{interp}\AgdaSpace{}%
\AgdaInductiveConstructor{Id}\AgdaSpace{}%
\AgdaBound{A}%
\>[28]\AgdaSymbol{=}\AgdaSpace{}%
\AgdaBound{A}\<%
\\
\>[8]\AgdaFunction{interp}\AgdaSpace{}%
\AgdaSymbol{(}\AgdaBound{f}\AgdaSpace{}%
\AgdaOperator{\AgdaInductiveConstructor{⊕}}\AgdaSpace{}%
\AgdaBound{g}\AgdaSymbol{)}\AgdaSpace{}%
\AgdaBound{A}%
\>[28]\AgdaSymbol{=}\AgdaSpace{}%
\AgdaFunction{interp}\AgdaSpace{}%
\AgdaBound{f}\AgdaSpace{}%
\AgdaBound{A}\AgdaSpace{}%
\AgdaOperator{\AgdaDatatype{+}}\AgdaSpace{}%
\AgdaFunction{interp}\AgdaSpace{}%
\AgdaBound{g}\AgdaSpace{}%
\AgdaBound{A}\<%
\\
\>[8]\AgdaFunction{interp}\AgdaSpace{}%
\AgdaSymbol{(}\AgdaBound{f}\AgdaSpace{}%
\AgdaOperator{\AgdaInductiveConstructor{⊗}}\AgdaSpace{}%
\AgdaBound{g}\AgdaSymbol{)}\AgdaSpace{}%
\AgdaBound{A}%
\>[28]\AgdaSymbol{=}\AgdaSpace{}%
\AgdaFunction{interp}\AgdaSpace{}%
\AgdaBound{f}\AgdaSpace{}%
\AgdaBound{A}\AgdaSpace{}%
\AgdaOperator{\AgdaDatatype{×}}\AgdaSpace{}%
\AgdaFunction{interp}\AgdaSpace{}%
\AgdaBound{g}\AgdaSpace{}%
\AgdaBound{A}\<%
\\
\\[\AgdaEmptyExtraSkip]%
\>[8]\AgdaKeyword{data}\AgdaSpace{}%
\AgdaDatatype{Fix}\AgdaSpace{}%
\AgdaSymbol{(}\AgdaBound{f}\AgdaSpace{}%
\AgdaSymbol{:}\AgdaSpace{}%
\AgdaDatatype{Poly}\AgdaSymbol{)}\AgdaSpace{}%
\AgdaSymbol{:}\AgdaSpace{}%
\AgdaPrimitive{Set}\AgdaSpace{}%
\AgdaKeyword{where}\<%
\\
\>[8][@{}l@{\AgdaIndent{0}}]%
\>[12]\AgdaInductiveConstructor{fix}\AgdaSpace{}%
\AgdaSymbol{:}\AgdaSpace{}%
\AgdaFunction{interp}\AgdaSpace{}%
\AgdaBound{f}\AgdaSpace{}%
\AgdaSymbol{(}\AgdaDatatype{Fix}\AgdaSpace{}%
\AgdaBound{f}\AgdaSymbol{)}\AgdaSpace{}%
\AgdaSymbol{→}\AgdaSpace{}%
\AgdaDatatype{Fix}\AgdaSpace{}%
\AgdaBound{f}\<%
\end{code}

One can implement a generic \AgdaFunction{cata} on \AgdaDatatype{Fix}.
This is left as an exercise for the reader. You can find the answer at \cite{tutorial}.

In the same spirit, \cite{gallais} described how to manipulate and reason about
generic syntax. In abstract nonsense, this can be recast as initial algebras
of functors in the aforementioned presheaf category.
But apart from polynomial functors, we get to use more functors related to
the binding structure. In particular, the weakening operation induces
a new functor that corresponds to \AgdaInductiveConstructor{lam}.

This way of describing things is cool, clean and intuitive.
Since we build up a universe of syntaxes inductively, we have absolute
control over them. And theorems etc. can be proven by induction on
the syntaxes, which is also convenient.
However, it is more suited to theorem proving than to practical programming:
For the same reason, we almost always prefer using inductive datatypes
to using \AgdaDatatype{Fix}, even though they are equivalent in expressivity!%
\footnote{No, they are not. But non-dependent inductive datatypes can
be represented with \AgdaDatatype{Fix}, once it is appropriately extended
to allow infinitary sums and products.}\footnote{Also, with the Univalence
Axiom, we can seamlessly transport all the theorems from there to our
homemade syntax (e.g. as defined in our first section), thus taking
the best part of both sides.}

Therefore, I shall now tread this road less taken. And let us see what
plight awaits us!

\section{The Road Less Taken}

Instead of building a universe on which we have absolute control, let's
choose to \emph{place restrictions} on a large existing universe so we have
\emph{minimum} control. Now, all we know is that we have a \AgdaFunction{Scope}.
What more can we say about it?

Well, let's do weakening first:

\AgdaTarget{Weakening, weaken}
\begin{code}%
\>[4]\AgdaKeyword{record}\AgdaSpace{}%
\AgdaRecord{Weakening}\AgdaSpace{}%
\AgdaSymbol{(}\AgdaBound{𝒞}\AgdaSpace{}%
\AgdaSymbol{:}\AgdaSpace{}%
\AgdaFunction{Scope}\AgdaSymbol{)}\AgdaSpace{}%
\AgdaSymbol{:}\AgdaSpace{}%
\AgdaPrimitive{Set}\AgdaSpace{}%
\AgdaKeyword{where}\<%
\\
\>[4][@{}l@{\AgdaIndent{0}}]%
\>[8]\AgdaKeyword{field}\<%
\\
\>[8][@{}l@{\AgdaIndent{0}}]%
\>[12]\AgdaField{weaken}\AgdaSpace{}%
\AgdaSymbol{:}%
\>[26]\AgdaSymbol{(}\AgdaFunction{𝓥}\AgdaSpace{}%
\AgdaOperator{\AgdaFunction{⇒}}\AgdaSpace{}%
\AgdaBound{𝒞}\AgdaSymbol{)}\AgdaSpace{}%
\AgdaGeneralizable{Γ}%
\>[44]\AgdaGeneralizable{Δ}\<%
\\
\>[12][@{}l@{\AgdaIndent{0}}]%
\>[16]\AgdaSymbol{→}\AgdaSpace{}%
\AgdaSymbol{∀}\AgdaSpace{}%
\AgdaBound{i}%
\>[23]\AgdaSymbol{→}%
\>[26]\AgdaSymbol{(}\AgdaFunction{𝓥}\AgdaSpace{}%
\AgdaOperator{\AgdaFunction{⇒}}\AgdaSpace{}%
\AgdaBound{𝒞}\AgdaSymbol{)}\AgdaSpace{}%
\AgdaSymbol{(}\AgdaGeneralizable{Γ}\AgdaSpace{}%
\AgdaOperator{\AgdaInductiveConstructor{◂}}\AgdaSpace{}%
\AgdaBound{i}\AgdaSymbol{)}%
\>[44]\AgdaSymbol{(}\AgdaGeneralizable{Δ}\AgdaSpace{}%
\AgdaOperator{\AgdaInductiveConstructor{◂}}\AgdaSpace{}%
\AgdaBound{i}\AgdaSymbol{)}\<%
\\
\>[4]\AgdaKeyword{open}\AgdaSpace{}%
\AgdaModule{Weakening}\AgdaSpace{}%
\AgdaSymbol{⦃...⦄}\AgdaSpace{}%
\AgdaKeyword{public}\<%
\end{code}

Note that the $⦃\dots⦄$ tells Agda to treat this record type similarly to
\emph{typeclasses} as in Haskell. We can declare instances of this typeclass,
which will be automatically used to infer arguments.

\begin{code}[hide]%
\>[4]\AgdaKeyword{infixl}\AgdaSpace{}%
\AgdaNumber{50}\AgdaSpace{}%
\AgdaOperator{\AgdaFunction{\AgdaUnderscore{}≪\AgdaUnderscore{}}}\<%
\\
\>[4]\AgdaOperator{\AgdaFunction{\AgdaUnderscore{}≪\AgdaUnderscore{}}}%
\>[614I]\AgdaSymbol{:}\AgdaSpace{}%
\AgdaSymbol{⦃}\AgdaSpace{}%
\AgdaRecord{Weakening}\AgdaSpace{}%
\AgdaGeneralizable{𝒞}\AgdaSpace{}%
\AgdaSymbol{⦄}\<%
\\
\>[.][@{}l@{}]\<[614I]%
\>[8]\AgdaSymbol{→}\AgdaSpace{}%
\AgdaSymbol{∀}\AgdaSpace{}%
\AgdaSymbol{\{}\AgdaBound{Γ}\AgdaSpace{}%
\AgdaBound{Δ}\AgdaSymbol{\}}%
\>[19]\AgdaSymbol{→}\AgdaSpace{}%
\AgdaSymbol{(}\AgdaFunction{𝓥}\AgdaSpace{}%
\AgdaOperator{\AgdaFunction{⇒}}\AgdaSpace{}%
\AgdaGeneralizable{𝒞}\AgdaSymbol{)}\AgdaSpace{}%
\AgdaBound{Γ}%
\>[38]\AgdaBound{Δ}\<%
\\
\>[8]\AgdaSymbol{→}\AgdaSpace{}%
\AgdaSymbol{∀}\AgdaSpace{}%
\AgdaBound{i}%
\>[19]\AgdaSymbol{→}\AgdaSpace{}%
\AgdaSymbol{(}\AgdaFunction{𝓥}\AgdaSpace{}%
\AgdaOperator{\AgdaFunction{⇒}}\AgdaSpace{}%
\AgdaGeneralizable{𝒞}\AgdaSymbol{)}\AgdaSpace{}%
\AgdaSymbol{(}\AgdaBound{Γ}\AgdaSpace{}%
\AgdaOperator{\AgdaInductiveConstructor{◂}}\AgdaSpace{}%
\AgdaBound{i}\AgdaSymbol{)}%
\>[38]\AgdaSymbol{(}\AgdaBound{Δ}\AgdaSpace{}%
\AgdaOperator{\AgdaInductiveConstructor{◂}}\AgdaSpace{}%
\AgdaBound{i}\AgdaSymbol{)}\<%
\end{code}

A convenient infix:
\AgdaTarget{\_≪\_}
\begin{code}%
\>[4]\AgdaOperator{\AgdaFunction{\AgdaUnderscore{}≪\AgdaUnderscore{}}}\AgdaSpace{}%
\AgdaSymbol{=}\AgdaSpace{}%
\AgdaField{weaken}\<%
\end{code}

Sanity check: \AgdaFunction{𝓥} itself can be weakened:
\begin{code}%
\>[4]\AgdaKeyword{instance}\<%
\\
\>[4][@{}l@{\AgdaIndent{0}}]%
\>[8]\AgdaFunction{𝓥ʷ}\AgdaSpace{}%
\AgdaSymbol{:}\AgdaSpace{}%
\AgdaRecord{Weakening}\AgdaSpace{}%
\AgdaFunction{𝓥}\<%
\\
\>[8]\AgdaFunction{𝓥ʷ}\AgdaSpace{}%
\AgdaSymbol{.}\AgdaField{weaken}\AgdaSpace{}%
\AgdaBound{ρ}\AgdaSpace{}%
\AgdaBound{i}\AgdaSpace{}%
\AgdaInductiveConstructor{𝕫}\AgdaSpace{}%
\AgdaSymbol{=}\AgdaSpace{}%
\AgdaInductiveConstructor{𝕫}\<%
\\
\>[8]\AgdaFunction{𝓥ʷ}\AgdaSpace{}%
\AgdaSymbol{.}\AgdaField{weaken}\AgdaSpace{}%
\AgdaBound{ρ}\AgdaSpace{}%
\AgdaBound{i}\AgdaSpace{}%
\AgdaSymbol{(}\AgdaOperator{\AgdaInductiveConstructor{𝕤}}\AgdaSpace{}%
\AgdaBound{v}\AgdaSymbol{)}\AgdaSpace{}%
\AgdaSymbol{=}\AgdaSpace{}%
\AgdaOperator{\AgdaInductiveConstructor{𝕤}}\AgdaSpace{}%
\AgdaSymbol{(}\AgdaBound{ρ}\AgdaSpace{}%
\AgdaBound{v}\AgdaSymbol{)}\<%
\end{code}

Good. Next, we can start to extract the common pattern in the
renaming and substitution process. If the code is rewritten with our
combinators, the signature of \AgdaFunction{rename} would be something
like
\begin{code}[hide]%
\>[4]\AgdaFunction{renameType}\AgdaSpace{}%
\AgdaSymbol{:}\AgdaSpace{}%
\AgdaFunction{Scope}\AgdaSpace{}%
\AgdaSymbol{→}\AgdaSpace{}%
\AgdaPrimitive{Set}\<%
\\
\>[4]\AgdaFunction{renameType}\AgdaSpace{}%
\AgdaBound{𝒞}\AgdaSpace{}%
\AgdaSymbol{=}\<%
\end{code}
\begin{code}[inline]%
\>[4][@{}l@{\AgdaIndent{1}}]%
\>[8]\AgdaOperator{\AgdaFunction{⟦}}\AgdaSpace{}%
\AgdaFunction{𝓥}\AgdaSpace{}%
\AgdaOperator{\AgdaFunction{⇒}}\AgdaSpace{}%
\AgdaFunction{𝓥}\AgdaSpace{}%
\AgdaOperator{\AgdaFunction{⇛}}\AgdaSpace{}%
\AgdaBound{𝒞}\AgdaSpace{}%
\AgdaOperator{\AgdaFunction{⇒}}\AgdaSpace{}%
\AgdaBound{𝒞}\AgdaSpace{}%
\AgdaOperator{\AgdaFunction{⟧}}\<%
\end{code}
. And \AgdaFunction{subst} would be of type
\begin{code}[hide]%
\>[4]\AgdaFunction{substType}\AgdaSpace{}%
\AgdaSymbol{:}\AgdaSpace{}%
\AgdaFunction{Scope}\AgdaSpace{}%
\AgdaSymbol{→}\AgdaSpace{}%
\AgdaPrimitive{Set}\<%
\\
\>[4]\AgdaFunction{substType}\AgdaSpace{}%
\AgdaBound{𝒞}\AgdaSpace{}%
\AgdaSymbol{=}\<%
\end{code}
\begin{code}[inline]%
\>[4][@{}l@{\AgdaIndent{1}}]%
\>[8]\AgdaOperator{\AgdaFunction{⟦}}\AgdaSpace{}%
\AgdaFunction{𝓥}\AgdaSpace{}%
\AgdaOperator{\AgdaFunction{⇒}}\AgdaSpace{}%
\AgdaBound{𝒞}\AgdaSpace{}%
\AgdaOperator{\AgdaFunction{⇛}}\AgdaSpace{}%
\AgdaBound{𝒞}\AgdaSpace{}%
\AgdaOperator{\AgdaFunction{⇒}}\AgdaSpace{}%
\AgdaBound{𝒞}\AgdaSpace{}%
\AgdaOperator{\AgdaFunction{⟧}}\<%
\end{code}
.

Interesting! So looking at the types only, we would naturally come
to the generalization
\begin{code}[hide]%
\>[4]\AgdaFunction{genType}\AgdaSpace{}%
\AgdaSymbol{:}\AgdaSpace{}%
\AgdaFunction{Scope}\AgdaSpace{}%
\AgdaSymbol{→}\AgdaSpace{}%
\AgdaFunction{Scope}\AgdaSpace{}%
\AgdaSymbol{→}\AgdaSpace{}%
\AgdaPrimitive{Set}\<%
\\
\>[4]\AgdaFunction{genType}\AgdaSpace{}%
\AgdaBound{𝒞}\AgdaSpace{}%
\AgdaBound{𝒜}\AgdaSpace{}%
\AgdaSymbol{=}\<%
\end{code}
\begin{code}[inline]%
\>[4][@{}l@{\AgdaIndent{1}}]%
\>[8]\AgdaOperator{\AgdaFunction{⟦}}\AgdaSpace{}%
\AgdaFunction{𝓥}\AgdaSpace{}%
\AgdaOperator{\AgdaFunction{⇒}}\AgdaSpace{}%
\AgdaBound{𝒜}\AgdaSpace{}%
\AgdaOperator{\AgdaFunction{⇛}}\AgdaSpace{}%
\AgdaBound{𝒞}\AgdaSpace{}%
\AgdaOperator{\AgdaFunction{⇒}}\AgdaSpace{}%
\AgdaBound{𝒞}\AgdaSpace{}%
\AgdaOperator{\AgdaFunction{⟧}}\<%
\end{code}
. Now of course, we can't define this for arbitrary \AgdaArgument{𝒜}.
So what more would we need? We have an assignment of variables to \AgdaArgument{𝒜},
and we are given an expression of \AgdaArgument{𝒞}. We need to replace
all the free variables in the expression according to the assignment.
So we definitely need a conversion
\begin{code}[hide]%
\>[4]\AgdaFunction{convType}\AgdaSpace{}%
\AgdaSymbol{:}\AgdaSpace{}%
\AgdaFunction{Scope}\AgdaSpace{}%
\AgdaSymbol{→}\AgdaSpace{}%
\AgdaFunction{Scope}\AgdaSpace{}%
\AgdaSymbol{→}\AgdaSpace{}%
\AgdaPrimitive{Set}\<%
\\
\>[4]\AgdaFunction{convType}\AgdaSpace{}%
\AgdaBound{𝒞}\AgdaSpace{}%
\AgdaBound{𝒜}\AgdaSpace{}%
\AgdaSymbol{=}\<%
\end{code}
\begin{code}[inline*]%
\>[4][@{}l@{\AgdaIndent{1}}]%
\>[8]\AgdaOperator{\AgdaFunction{[}}\AgdaSpace{}%
\AgdaBound{𝒜}\AgdaSpace{}%
\AgdaOperator{\AgdaFunction{⇒}}\AgdaSpace{}%
\AgdaBound{𝒞}\AgdaSpace{}%
\AgdaOperator{\AgdaFunction{]}}\<%
\end{code}
from the assigned \AgdaArgument{𝒜}'s to \AgdaArgument{𝒞}'s.

During the process, we also need to be able to push into binders.
Therefore we also need to ``weaken''. This brings us to the complete
type signature:
\begin{code}[hide]%
\>[4]\AgdaFunction{mapType}\AgdaSpace{}%
\AgdaSymbol{:}\AgdaSpace{}%
\AgdaFunction{Scope}\AgdaSpace{}%
\AgdaSymbol{→}\AgdaSpace{}%
\AgdaFunction{Scope}\AgdaSpace{}%
\AgdaSymbol{→}\AgdaSpace{}%
\AgdaPrimitive{Set}\<%
\\
\>[4]\AgdaFunction{mapType}%
\>[720I]\AgdaBound{𝒞}\AgdaSpace{}%
\AgdaBound{𝒜}\AgdaSpace{}%
\AgdaSymbol{=}\<%
\end{code}
\begin{code}%
\>[.][@{}l@{}]\<[720I]%
\>[12]\AgdaSymbol{⦃}\AgdaSpace{}%
\AgdaRecord{Weakening}\AgdaSpace{}%
\AgdaBound{𝒜}\AgdaSpace{}%
\AgdaSymbol{⦄}\<%
\\
\>[4][@{}l@{\AgdaIndent{1}}]%
\>[8]\AgdaSymbol{→}%
\>[12]\AgdaOperator{\AgdaFunction{[}}\AgdaSpace{}%
\AgdaBound{𝒜}\AgdaSpace{}%
\AgdaOperator{\AgdaFunction{⇒}}\AgdaSpace{}%
\AgdaBound{𝒞}\AgdaSpace{}%
\AgdaOperator{\AgdaFunction{]}}\<%
\\
\>[8]\AgdaSymbol{→}%
\>[12]\AgdaOperator{\AgdaFunction{⟦}}\AgdaSpace{}%
\AgdaFunction{𝓥}\AgdaSpace{}%
\AgdaOperator{\AgdaFunction{⇒}}\AgdaSpace{}%
\AgdaBound{𝒜}\AgdaSpace{}%
\AgdaOperator{\AgdaFunction{⇛}}\AgdaSpace{}%
\AgdaBound{𝒞}\AgdaSpace{}%
\AgdaOperator{\AgdaFunction{⇒}}\AgdaSpace{}%
\AgdaBound{𝒞}\AgdaSpace{}%
\AgdaOperator{\AgdaFunction{⟧}}\<%
\end{code}

Last but not least, we need to ensure that every variable is also
a legitimate term, otherwise we wouldn't have much to work with. Packing
all of these up gives us another typeclass:

\AgdaTarget{Syntax, var, map}
\begin{code}%
\>[4]\AgdaKeyword{record}\AgdaSpace{}%
\AgdaRecord{Syntax}\AgdaSpace{}%
\AgdaSymbol{(}\AgdaBound{𝒞}\AgdaSpace{}%
\AgdaSymbol{:}\AgdaSpace{}%
\AgdaFunction{Scope}\AgdaSymbol{)}\AgdaSpace{}%
\AgdaSymbol{:}\AgdaSpace{}%
\AgdaPrimitive{Set₁}\AgdaSpace{}%
\AgdaKeyword{where}\<%
\\
\>[4][@{}l@{\AgdaIndent{0}}]%
\>[8]\AgdaKeyword{field}\<%
\\
\>[8][@{}l@{\AgdaIndent{0}}]%
\>[12]\AgdaField{var}%
\>[17]\AgdaSymbol{:}\AgdaSpace{}%
\AgdaOperator{\AgdaFunction{[}}\AgdaSpace{}%
\AgdaFunction{𝓥}\AgdaSpace{}%
\AgdaOperator{\AgdaFunction{⇒}}\AgdaSpace{}%
\AgdaBound{𝒞}\AgdaSpace{}%
\AgdaOperator{\AgdaFunction{]}}\<%
\\
\>[12]\AgdaField{map}%
\>[17]\AgdaSymbol{:}%
\>[20]\AgdaSymbol{⦃}\AgdaSpace{}%
\AgdaRecord{Weakening}\AgdaSpace{}%
\AgdaGeneralizable{𝒜}\AgdaSpace{}%
\AgdaSymbol{⦄}\<%
\\
\>[12][@{}l@{\AgdaIndent{0}}]%
\>[16]\AgdaSymbol{→}%
\>[20]\AgdaOperator{\AgdaFunction{[}}\AgdaSpace{}%
\AgdaGeneralizable{𝒜}\AgdaSpace{}%
\AgdaOperator{\AgdaFunction{⇒}}\AgdaSpace{}%
\AgdaBound{𝒞}\AgdaSpace{}%
\AgdaOperator{\AgdaFunction{]}}\<%
\\
\>[16]\AgdaSymbol{→}%
\>[20]\AgdaOperator{\AgdaFunction{⟦}}\AgdaSpace{}%
\AgdaFunction{𝓥}\AgdaSpace{}%
\AgdaOperator{\AgdaFunction{⇒}}\AgdaSpace{}%
\AgdaGeneralizable{𝒜}\AgdaSpace{}%
\AgdaOperator{\AgdaFunction{⇛}}\AgdaSpace{}%
\AgdaBound{𝒞}\AgdaSpace{}%
\AgdaOperator{\AgdaFunction{⇒}}\AgdaSpace{}%
\AgdaBound{𝒞}\AgdaSpace{}%
\AgdaOperator{\AgdaFunction{⟧}}\<%
\end{code}

Now we can implement renaming and substitution based on the typeclass
methods, and when we are using them in practice, we only need to provide
the implementation of \AgdaField{var} and \AgdaField{map}. Sweet!

Renaming is a piece of cake:

\AgdaTarget{rename}
\begin{code}%
\>[8]\AgdaFunction{rename}\AgdaSpace{}%
\AgdaSymbol{:}\AgdaSpace{}%
\AgdaOperator{\AgdaFunction{⟦}}\AgdaSpace{}%
\AgdaFunction{𝓥}\AgdaSpace{}%
\AgdaOperator{\AgdaFunction{⇒}}\AgdaSpace{}%
\AgdaFunction{𝓥}\AgdaSpace{}%
\AgdaOperator{\AgdaFunction{⇛}}\AgdaSpace{}%
\AgdaBound{𝒞}\AgdaSpace{}%
\AgdaOperator{\AgdaFunction{⇒}}\AgdaSpace{}%
\AgdaBound{𝒞}\AgdaSpace{}%
\AgdaOperator{\AgdaFunction{⟧}}\<%
\\
\>[8]\AgdaFunction{rename}\AgdaSpace{}%
\AgdaSymbol{=}\AgdaSpace{}%
\AgdaField{map}\AgdaSpace{}%
\AgdaField{var}\<%
\end{code}

Note that since we already told Agda that \AgdaFunction{𝓥} has
\AgdaField{weaken}ing, we don't need to mention that at all here.

Next, before we start to implement substitution, we need:

\begin{code}%
\>[8]\AgdaFunction{Syntaxʷ}\AgdaSpace{}%
\AgdaSymbol{:}\AgdaSpace{}%
\AgdaRecord{Weakening}\AgdaSpace{}%
\AgdaBound{𝒞}\<%
\\
\>[8]\AgdaFunction{Syntaxʷ}\AgdaSpace{}%
\AgdaSymbol{.}\AgdaField{weaken}\AgdaSpace{}%
\AgdaBound{σ}\AgdaSpace{}%
\AgdaBound{i}\AgdaSpace{}%
\AgdaInductiveConstructor{𝕫}\AgdaSpace{}%
\AgdaSymbol{=}\AgdaSpace{}%
\AgdaField{var}\AgdaSpace{}%
\AgdaInductiveConstructor{𝕫}\<%
\\
\>[8]\AgdaFunction{Syntaxʷ}\AgdaSpace{}%
\AgdaSymbol{.}\AgdaField{weaken}\AgdaSpace{}%
\AgdaBound{σ}\AgdaSpace{}%
\AgdaBound{i}\AgdaSpace{}%
\AgdaSymbol{(}\AgdaOperator{\AgdaInductiveConstructor{𝕤}}\AgdaSpace{}%
\AgdaBound{v}\AgdaSymbol{)}\AgdaSpace{}%
\AgdaSymbol{=}\AgdaSpace{}%
\AgdaFunction{rename}\AgdaSpace{}%
\AgdaOperator{\AgdaInductiveConstructor{𝕤\AgdaUnderscore{}}}\AgdaSpace{}%
\AgdaSymbol{(}\AgdaBound{σ}\AgdaSpace{}%
\AgdaBound{v}\AgdaSymbol{)}\<%
\end{code}

For technical reasons, it is not very convenient for us to make it an
instance. So for substitution, we will tell Agda about \AgdaFunction{Weakening}
manually.

\AgdaTarget{subst}
\begin{code}%
\>[8]\AgdaFunction{subst}\AgdaSpace{}%
\AgdaSymbol{:}\AgdaSpace{}%
\AgdaOperator{\AgdaFunction{⟦}}\AgdaSpace{}%
\AgdaFunction{𝓥}\AgdaSpace{}%
\AgdaOperator{\AgdaFunction{⇒}}\AgdaSpace{}%
\AgdaBound{𝒞}\AgdaSpace{}%
\AgdaOperator{\AgdaFunction{⇛}}\AgdaSpace{}%
\AgdaBound{𝒞}\AgdaSpace{}%
\AgdaOperator{\AgdaFunction{⇒}}\AgdaSpace{}%
\AgdaBound{𝒞}\AgdaSpace{}%
\AgdaOperator{\AgdaFunction{⟧}}\<%
\\
\>[8]\AgdaFunction{subst}\AgdaSpace{}%
\AgdaSymbol{=}\AgdaSpace{}%
\AgdaField{map}\AgdaSpace{}%
\AgdaFunction{id}\<%
\\
\>[8][@{}l@{\AgdaIndent{0}}]%
\>[12]\AgdaKeyword{where}\AgdaSpace{}%
\AgdaKeyword{instance}\AgdaSpace{}%
\AgdaFunction{\AgdaUnderscore{}}\AgdaSpace{}%
\AgdaSymbol{=}\AgdaSpace{}%
\AgdaFunction{Syntaxʷ}\<%
\end{code}

And there you have it. Finally, since we often need to substitute
only one variable (the rightmost one in the context), we make a little
helper function:

\begin{code}%
\>[8]\AgdaOperator{\AgdaFunction{𝕫/\AgdaUnderscore{}}}\AgdaSpace{}%
\AgdaSymbol{:}\AgdaSpace{}%
\AgdaBound{𝒞}\AgdaSpace{}%
\AgdaGeneralizable{Γ}\AgdaSpace{}%
\AgdaGeneralizable{i}\AgdaSpace{}%
\AgdaSymbol{→}\AgdaSpace{}%
\AgdaSymbol{(}\AgdaFunction{𝓥}\AgdaSpace{}%
\AgdaOperator{\AgdaFunction{⇒}}\AgdaSpace{}%
\AgdaBound{𝒞}\AgdaSymbol{)}\AgdaSpace{}%
\AgdaSymbol{(}\AgdaGeneralizable{Γ}\AgdaSpace{}%
\AgdaOperator{\AgdaInductiveConstructor{◂}}\AgdaSpace{}%
\AgdaGeneralizable{i}\AgdaSymbol{)}\AgdaSpace{}%
\AgdaGeneralizable{Γ}\<%
\\
\>[8]\AgdaSymbol{(}\AgdaOperator{\AgdaFunction{𝕫/}}\AgdaSpace{}%
\AgdaBound{t}\AgdaSymbol{)}\AgdaSpace{}%
\AgdaInductiveConstructor{𝕫}\AgdaSpace{}%
\AgdaSymbol{=}\AgdaSpace{}%
\AgdaBound{t}\<%
\\
\>[8]\AgdaSymbol{(}\AgdaOperator{\AgdaFunction{𝕫/}}\AgdaSpace{}%
\AgdaBound{t}\AgdaSymbol{)}\AgdaSpace{}%
\AgdaSymbol{(}\AgdaOperator{\AgdaInductiveConstructor{𝕤}}\AgdaSpace{}%
\AgdaBound{v}\AgdaSymbol{)}\AgdaSpace{}%
\AgdaSymbol{=}\AgdaSpace{}%
\AgdaField{var}\AgdaSpace{}%
\AgdaBound{v}\<%
\\
\>[8]\AgdaKeyword{infixr}\AgdaSpace{}%
\AgdaNumber{6}\AgdaSpace{}%
\AgdaOperator{\AgdaFunction{𝕫/\AgdaUnderscore{}}}\<%
\\
\>[4]\AgdaKeyword{open}\AgdaSpace{}%
\AgdaModule{Syntax}\AgdaSpace{}%
\AgdaSymbol{⦃...⦄}\AgdaSpace{}%
\AgdaKeyword{public}\<%
\end{code}

Let's see how things goes if we rewrite the simply typed $\lambda$-calculus
with the tools we just developed. First, we need to show that
\AgdaFunction{𝓥} itself is an instance of \AgdaFunction{Syntax}:

\begin{code}%
\>[4]\AgdaKeyword{instance}\<%
\\
\>[4][@{}l@{\AgdaIndent{0}}]%
\>[8]\AgdaFunction{𝓥ˢ}\AgdaSpace{}%
\AgdaSymbol{:}\AgdaSpace{}%
\AgdaRecord{Syntax}\AgdaSpace{}%
\AgdaFunction{𝓥}\<%
\\
\>[8]\AgdaFunction{𝓥ˢ}\AgdaSpace{}%
\AgdaSymbol{.}\AgdaField{var}\AgdaSpace{}%
\AgdaSymbol{=}\AgdaSpace{}%
\AgdaFunction{id}\<%
\\
\>[8]\AgdaFunction{𝓥ˢ}\AgdaSpace{}%
\AgdaSymbol{.}\AgdaField{map}\AgdaSpace{}%
\AgdaBound{𝑓}\AgdaSpace{}%
\AgdaBound{σ}\AgdaSpace{}%
\AgdaBound{v}\AgdaSpace{}%
\AgdaSymbol{=}\AgdaSpace{}%
\AgdaBound{𝑓}\AgdaSpace{}%
\AgdaSymbol{(}\AgdaBound{σ}\AgdaSpace{}%
\AgdaBound{v}\AgdaSymbol{)}\<%
\end{code}
\begin{code}[hide]%
\>[0]\AgdaKeyword{module}\AgdaSpace{}%
\AgdaModule{TestLambda}\AgdaSpace{}%
\AgdaKeyword{where}\<%
\\
\>[0][@{}l@{\AgdaIndent{0}}]%
\>[4]\AgdaKeyword{data}\AgdaSpace{}%
\AgdaDatatype{𝕋}\AgdaSpace{}%
\AgdaSymbol{:}\AgdaSpace{}%
\AgdaPrimitive{Set}\AgdaSpace{}%
\AgdaKeyword{where}\<%
\\
\>[4][@{}l@{\AgdaIndent{0}}]%
\>[8]\AgdaInductiveConstructor{ι}%
\>[13]\AgdaSymbol{:}\AgdaSpace{}%
\AgdaDatatype{𝕋}\<%
\\
\>[8]\AgdaOperator{\AgdaInductiveConstructor{\AgdaUnderscore{}⟶\AgdaUnderscore{}}}%
\>[13]\AgdaSymbol{:}\AgdaSpace{}%
\AgdaDatatype{𝕋}\AgdaSpace{}%
\AgdaSymbol{→}\AgdaSpace{}%
\AgdaDatatype{𝕋}\AgdaSpace{}%
\AgdaSymbol{→}\AgdaSpace{}%
\AgdaDatatype{𝕋}\<%
\\
\>[4]\AgdaKeyword{infixr}\AgdaSpace{}%
\AgdaNumber{15}\AgdaSpace{}%
\AgdaOperator{\AgdaInductiveConstructor{\AgdaUnderscore{}⟶\AgdaUnderscore{}}}\<%
\\
\>[4]\AgdaKeyword{private}\AgdaSpace{}%
\AgdaKeyword{variable}\<%
\\
\>[4][@{}l@{\AgdaIndent{0}}]%
\>[8]\AgdaGeneralizable{i}\AgdaSpace{}%
\AgdaGeneralizable{j}\AgdaSpace{}%
\AgdaSymbol{:}\AgdaSpace{}%
\AgdaDatatype{𝕋}\<%
\\
\>[8]\AgdaGeneralizable{Γ}\AgdaSpace{}%
\AgdaGeneralizable{Δ}\AgdaSpace{}%
\AgdaGeneralizable{Θ}\AgdaSpace{}%
\AgdaSymbol{:}\AgdaSpace{}%
\AgdaDatatype{List}\AgdaSpace{}%
\AgdaDatatype{𝕋}\<%
\\
\>[8]\AgdaGeneralizable{𝒜}\AgdaSpace{}%
\AgdaGeneralizable{ℬ}\AgdaSpace{}%
\AgdaGeneralizable{𝒞}\AgdaSpace{}%
\AgdaGeneralizable{𝒟}\AgdaSpace{}%
\AgdaSymbol{:}\AgdaSpace{}%
\AgdaDatatype{List}\AgdaSpace{}%
\AgdaDatatype{𝕋}\AgdaSpace{}%
\AgdaSymbol{->}\AgdaSpace{}%
\AgdaDatatype{𝕋}\AgdaSpace{}%
\AgdaSymbol{->}\AgdaSpace{}%
\AgdaPrimitive{Set}\<%
\end{code}

Recall that we worked in a parameterized module.
Now we instantiate the parameter with the concrete type \AgdaDatatype{𝕋}.

\begin{code}%
\>[4]\AgdaKeyword{open}\AgdaSpace{}%
\AgdaModule{Abstraction}\AgdaSpace{}%
\AgdaDatatype{𝕋}\<%
\\
\>[4]\AgdaKeyword{infix}\AgdaSpace{}%
\AgdaNumber{5}\AgdaSpace{}%
\AgdaOperator{\AgdaDatatype{\AgdaUnderscore{}⊢\AgdaUnderscore{}}}\<%
\\
\>[4]\AgdaKeyword{data}\AgdaSpace{}%
\AgdaOperator{\AgdaDatatype{\AgdaUnderscore{}⊢\AgdaUnderscore{}}}\AgdaSpace{}%
\AgdaSymbol{:}\AgdaSpace{}%
\AgdaFunction{Scope}\AgdaSpace{}%
\AgdaKeyword{where}\<%
\\
\>[4][@{}l@{\AgdaIndent{0}}]%
\>[8]\AgdaOperator{\AgdaInductiveConstructor{𝕧\AgdaUnderscore{}}}%
\>[13]\AgdaSymbol{:}%
\>[20]\AgdaGeneralizable{Γ}%
\>[23]\AgdaOperator{\AgdaDatatype{∋}}\AgdaSpace{}%
\AgdaGeneralizable{i}%
\>[44]\AgdaSymbol{→}\AgdaSpace{}%
\AgdaGeneralizable{Γ}\AgdaSpace{}%
\AgdaOperator{\AgdaDatatype{⊢}}\AgdaSpace{}%
\AgdaGeneralizable{i}\<%
\\
\>[8]\AgdaOperator{\AgdaInductiveConstructor{\AgdaUnderscore{}∙\AgdaUnderscore{}}}%
\>[13]\AgdaSymbol{:}%
\>[20]\AgdaGeneralizable{Γ}%
\>[23]\AgdaOperator{\AgdaDatatype{⊢}}\AgdaSpace{}%
\AgdaGeneralizable{i}\AgdaSpace{}%
\AgdaOperator{\AgdaInductiveConstructor{⟶}}\AgdaSpace{}%
\AgdaGeneralizable{j}%
\>[35]\AgdaSymbol{→}\AgdaSpace{}%
\AgdaGeneralizable{Γ}\AgdaSpace{}%
\AgdaOperator{\AgdaDatatype{⊢}}\AgdaSpace{}%
\AgdaGeneralizable{i}%
\>[44]\AgdaSymbol{→}\AgdaSpace{}%
\AgdaGeneralizable{Γ}\AgdaSpace{}%
\AgdaOperator{\AgdaDatatype{⊢}}\AgdaSpace{}%
\AgdaGeneralizable{j}\<%
\\
\>[8]\AgdaOperator{\AgdaInductiveConstructor{ƛ\AgdaUnderscore{}}}%
\>[13]\AgdaSymbol{:}\AgdaSpace{}%
\AgdaGeneralizable{Γ}\AgdaSpace{}%
\AgdaOperator{\AgdaInductiveConstructor{◂}}%
\>[20]\AgdaGeneralizable{i}%
\>[23]\AgdaOperator{\AgdaDatatype{⊢}}\AgdaSpace{}%
\AgdaGeneralizable{j}%
\>[44]\AgdaSymbol{→}\AgdaSpace{}%
\AgdaGeneralizable{Γ}\AgdaSpace{}%
\AgdaOperator{\AgdaDatatype{⊢}}\AgdaSpace{}%
\AgdaGeneralizable{i}\AgdaSpace{}%
\AgdaOperator{\AgdaInductiveConstructor{⟶}}\AgdaSpace{}%
\AgdaGeneralizable{j}\<%
\\
\>[4]\AgdaKeyword{infixl}\AgdaSpace{}%
\AgdaNumber{20}\AgdaSpace{}%
\AgdaOperator{\AgdaInductiveConstructor{\AgdaUnderscore{}∙\AgdaUnderscore{}}}\<%
\\
\>[4]\AgdaKeyword{infixr}\AgdaSpace{}%
\AgdaNumber{10}\AgdaSpace{}%
\AgdaOperator{\AgdaInductiveConstructor{ƛ\AgdaUnderscore{}}}\<%
\\
\>[4]\AgdaKeyword{infixr}\AgdaSpace{}%
\AgdaNumber{100}\AgdaSpace{}%
\AgdaOperator{\AgdaInductiveConstructor{𝕧\AgdaUnderscore{}}}\<%
\\
\\[\AgdaEmptyExtraSkip]%
\>[4]\AgdaFunction{𝓣}\AgdaSpace{}%
\AgdaSymbol{=}\AgdaSpace{}%
\AgdaOperator{\AgdaDatatype{\AgdaUnderscore{}⊢\AgdaUnderscore{}}}\<%
\end{code}

Again, we used $\lambda$ for both Agda functions and $\lambda$-calculus
functions. Now we give the implementation for the \AgdaFunction{Syntax}
structure:
\begin{code}%
\>[4]\AgdaKeyword{instance}\<%
\\
\>[4][@{}l@{\AgdaIndent{0}}]%
\>[8]\AgdaFunction{𝓣ˢ}\AgdaSpace{}%
\AgdaSymbol{:}\AgdaSpace{}%
\AgdaRecord{Syntax}\AgdaSpace{}%
\AgdaFunction{𝓣}\<%
\\
\>[8]\AgdaFunction{𝓣ˢ}\AgdaSpace{}%
\AgdaSymbol{.}\AgdaField{var}\AgdaSpace{}%
\AgdaSymbol{=}\AgdaSpace{}%
\AgdaOperator{\AgdaInductiveConstructor{𝕧\AgdaUnderscore{}}}\<%
\end{code}

The \AgdaField{var} case is easy. What about \AgdaField{map}?

\begin{code}%
\>[8]\AgdaFunction{𝓣ˢ}\AgdaSpace{}%
\AgdaSymbol{.}\AgdaField{map}\AgdaSpace{}%
\AgdaSymbol{=}\AgdaSpace{}%
\AgdaFunction{helper}\<%
\\
\>[8][@{}l@{\AgdaIndent{0}}]%
\>[9]\AgdaKeyword{where}\<%
\\
\>[9][@{}l@{\AgdaIndent{0}}]%
\>[13]\AgdaFunction{helper}\AgdaSpace{}%
\AgdaSymbol{:}\AgdaSpace{}%
\AgdaSymbol{⦃}\AgdaSpace{}%
\AgdaRecord{Weakening}\AgdaSpace{}%
\AgdaGeneralizable{𝒜}\AgdaSpace{}%
\AgdaSymbol{⦄}\<%
\\
\>[13][@{}l@{\AgdaIndent{0}}]%
\>[17]\AgdaSymbol{→}\AgdaSpace{}%
\AgdaOperator{\AgdaFunction{[}}\AgdaSpace{}%
\AgdaGeneralizable{𝒜}\AgdaSpace{}%
\AgdaOperator{\AgdaFunction{⇒}}\AgdaSpace{}%
\AgdaFunction{𝓣}\AgdaSpace{}%
\AgdaOperator{\AgdaFunction{]}}\<%
\\
\>[17]\AgdaSymbol{→}\AgdaSpace{}%
\AgdaOperator{\AgdaFunction{⟦}}\AgdaSpace{}%
\AgdaFunction{𝓥}\AgdaSpace{}%
\AgdaOperator{\AgdaFunction{⇒}}\AgdaSpace{}%
\AgdaGeneralizable{𝒜}\AgdaSpace{}%
\AgdaOperator{\AgdaFunction{⇛}}\AgdaSpace{}%
\AgdaFunction{𝓣}\AgdaSpace{}%
\AgdaOperator{\AgdaFunction{⇒}}\AgdaSpace{}%
\AgdaFunction{𝓣}\AgdaSpace{}%
\AgdaOperator{\AgdaFunction{⟧}}\<%
\end{code}

Agda helps us fill in the goals pretty easily:
\begin{code}%
\>[13]\AgdaFunction{helper}\AgdaSpace{}%
\AgdaBound{𝑓}\AgdaSpace{}%
\AgdaBound{σ}\AgdaSpace{}%
\AgdaSymbol{(}\AgdaOperator{\AgdaInductiveConstructor{𝕧}}\AgdaSpace{}%
\AgdaBound{v}\AgdaSymbol{)}%
\>[33]\AgdaSymbol{=}\AgdaSpace{}%
\AgdaBound{𝑓}\AgdaSpace{}%
\AgdaSymbol{(}\AgdaBound{σ}\AgdaSpace{}%
\AgdaBound{v}\AgdaSymbol{)}\<%
\\
\>[13]\AgdaFunction{helper}\AgdaSpace{}%
\AgdaBound{𝑓}\AgdaSpace{}%
\AgdaBound{σ}\AgdaSpace{}%
\AgdaSymbol{(}\AgdaBound{t}\AgdaSpace{}%
\AgdaOperator{\AgdaInductiveConstructor{∙}}\AgdaSpace{}%
\AgdaBound{s}\AgdaSymbol{)}%
\>[33]\AgdaSymbol{=}\AgdaSpace{}%
\AgdaFunction{helper}\AgdaSpace{}%
\AgdaBound{𝑓}\AgdaSpace{}%
\AgdaBound{σ}\AgdaSpace{}%
\AgdaBound{t}\AgdaSpace{}%
\AgdaOperator{\AgdaInductiveConstructor{∙}}\AgdaSpace{}%
\AgdaFunction{helper}\AgdaSpace{}%
\AgdaBound{𝑓}\AgdaSpace{}%
\AgdaBound{σ}\AgdaSpace{}%
\AgdaBound{s}\<%
\\
\>[13]\AgdaFunction{helper}\AgdaSpace{}%
\AgdaBound{𝑓}\AgdaSpace{}%
\AgdaBound{σ}\AgdaSpace{}%
\AgdaSymbol{(}\AgdaOperator{\AgdaInductiveConstructor{ƛ}}\AgdaSpace{}%
\AgdaBound{t}\AgdaSymbol{)}%
\>[33]\AgdaSymbol{=}\AgdaSpace{}%
\AgdaOperator{\AgdaInductiveConstructor{ƛ}}\AgdaSpace{}%
\AgdaFunction{helper}\AgdaSpace{}%
\AgdaBound{𝑓}\AgdaSpace{}%
\AgdaSymbol{(}\AgdaBound{σ}\AgdaSpace{}%
\AgdaOperator{\AgdaFunction{≪}}\AgdaSpace{}%
\AgdaSymbol{\AgdaUnderscore{})}\AgdaSpace{}%
\AgdaBound{t}\<%
\end{code}

We now get substitution for free! For example:
\begin{code}%
\>[4]\AgdaFunction{A}\AgdaSpace{}%
\AgdaSymbol{:}\AgdaSpace{}%
\AgdaInductiveConstructor{∅}\AgdaSpace{}%
\AgdaOperator{\AgdaInductiveConstructor{◂}}\AgdaSpace{}%
\AgdaInductiveConstructor{ι}\AgdaSpace{}%
\AgdaOperator{\AgdaInductiveConstructor{⟶}}\AgdaSpace{}%
\AgdaInductiveConstructor{ι}\AgdaSpace{}%
\AgdaOperator{\AgdaDatatype{⊢}}\AgdaSpace{}%
\AgdaInductiveConstructor{ι}\AgdaSpace{}%
\AgdaOperator{\AgdaInductiveConstructor{⟶}}\AgdaSpace{}%
\AgdaInductiveConstructor{ι}\<%
\\
\>[4]\AgdaFunction{A}\AgdaSpace{}%
\AgdaSymbol{=}\AgdaSpace{}%
\AgdaOperator{\AgdaInductiveConstructor{ƛ}}\AgdaSpace{}%
\AgdaOperator{\AgdaInductiveConstructor{𝕧}}\AgdaSpace{}%
\AgdaOperator{\AgdaInductiveConstructor{𝕤}}\AgdaSpace{}%
\AgdaInductiveConstructor{𝕫}\AgdaSpace{}%
\AgdaOperator{\AgdaInductiveConstructor{∙}}\AgdaSpace{}%
\AgdaSymbol{(}\AgdaOperator{\AgdaInductiveConstructor{𝕧}}\AgdaSpace{}%
\AgdaOperator{\AgdaInductiveConstructor{𝕤}}\AgdaSpace{}%
\AgdaInductiveConstructor{𝕫}\AgdaSpace{}%
\AgdaOperator{\AgdaInductiveConstructor{∙}}\AgdaSpace{}%
\AgdaOperator{\AgdaInductiveConstructor{𝕧}}\AgdaSpace{}%
\AgdaInductiveConstructor{𝕫}\AgdaSymbol{)}\<%
\\
\\[\AgdaEmptyExtraSkip]%
\>[4]\AgdaFunction{B}\AgdaSpace{}%
\AgdaSymbol{:}\AgdaSpace{}%
\AgdaInductiveConstructor{∅}\AgdaSpace{}%
\AgdaOperator{\AgdaDatatype{⊢}}\AgdaSpace{}%
\AgdaInductiveConstructor{ι}\AgdaSpace{}%
\AgdaOperator{\AgdaInductiveConstructor{⟶}}\AgdaSpace{}%
\AgdaInductiveConstructor{ι}\<%
\\
\>[4]\AgdaFunction{B}\AgdaSpace{}%
\AgdaSymbol{=}\AgdaSpace{}%
\AgdaOperator{\AgdaInductiveConstructor{ƛ}}\AgdaSpace{}%
\AgdaOperator{\AgdaInductiveConstructor{𝕧}}\AgdaSpace{}%
\AgdaInductiveConstructor{𝕫}\<%
\\
\\[\AgdaEmptyExtraSkip]%
\>[4]\AgdaFunction{A[B]}\AgdaSpace{}%
\AgdaSymbol{=}\AgdaSpace{}%
\AgdaFunction{subst}\AgdaSpace{}%
\AgdaSymbol{(}\AgdaOperator{\AgdaFunction{𝕫/}}\AgdaSpace{}%
\AgdaFunction{B}\AgdaSymbol{)}\AgdaSpace{}%
\AgdaFunction{A}\<%
\\
\>[4]\AgdaFunction{\AgdaUnderscore{}}\AgdaSpace{}%
\AgdaSymbol{:}\AgdaSpace{}%
\AgdaFunction{A[B]}\AgdaSpace{}%
\AgdaOperator{\AgdaDatatype{≡}}\AgdaSpace{}%
\AgdaOperator{\AgdaInductiveConstructor{ƛ}}\AgdaSpace{}%
\AgdaSymbol{(}\AgdaOperator{\AgdaInductiveConstructor{ƛ}}\AgdaSpace{}%
\AgdaOperator{\AgdaInductiveConstructor{𝕧}}\AgdaSpace{}%
\AgdaInductiveConstructor{𝕫}\AgdaSymbol{)}\AgdaSpace{}%
\AgdaOperator{\AgdaInductiveConstructor{∙}}\AgdaSpace{}%
\AgdaSymbol{((}\AgdaOperator{\AgdaInductiveConstructor{ƛ}}\AgdaSpace{}%
\AgdaOperator{\AgdaInductiveConstructor{𝕧}}\AgdaSpace{}%
\AgdaInductiveConstructor{𝕫}\AgdaSymbol{)}\AgdaSpace{}%
\AgdaOperator{\AgdaInductiveConstructor{∙}}\AgdaSpace{}%
\AgdaOperator{\AgdaInductiveConstructor{𝕧}}\AgdaSpace{}%
\AgdaInductiveConstructor{𝕫}\AgdaSymbol{)}\<%
\\
\>[4]\AgdaSymbol{\AgdaUnderscore{}}\AgdaSpace{}%
\AgdaSymbol{=}\AgdaSpace{}%
\AgdaInductiveConstructor{refl}\<%
\end{code}

\section{Higher-order Homomorphisms}

\begin{code}%
\>[0]\AgdaKeyword{module}\AgdaSpace{}%
\AgdaModule{HoHom}\AgdaSpace{}%
\AgdaSymbol{(}\AgdaBound{I}\AgdaSpace{}%
\AgdaSymbol{:}\AgdaSpace{}%
\AgdaPrimitive{Set}\AgdaSymbol{)}\AgdaSpace{}%
\AgdaKeyword{where}\<%
\end{code}
\begin{code}[hide]%
\>[0][@{}l@{\AgdaIndent{1}}]%
\>[4]\AgdaKeyword{open}\AgdaSpace{}%
\AgdaModule{Abstraction}\AgdaSpace{}%
\AgdaBound{I}\<%
\\
\>[4]\AgdaKeyword{private}\AgdaSpace{}%
\AgdaKeyword{variable}\<%
\\
\>[4][@{}l@{\AgdaIndent{0}}]%
\>[8]\AgdaGeneralizable{Γ}\AgdaSpace{}%
\AgdaGeneralizable{Δ}\AgdaSpace{}%
\AgdaGeneralizable{Θ}\AgdaSpace{}%
\AgdaGeneralizable{Ξ}\AgdaSpace{}%
\AgdaSymbol{:}\AgdaSpace{}%
\AgdaDatatype{List}\AgdaSpace{}%
\AgdaBound{I}\<%
\\
\>[8]\AgdaGeneralizable{i}\AgdaSpace{}%
\AgdaGeneralizable{j}\AgdaSpace{}%
\AgdaGeneralizable{k}\AgdaSpace{}%
\AgdaGeneralizable{l}\AgdaSpace{}%
\AgdaSymbol{:}\AgdaSpace{}%
\AgdaBound{I}\<%
\\
\>[8]\AgdaGeneralizable{𝒜}\AgdaSpace{}%
\AgdaGeneralizable{ℬ}\AgdaSpace{}%
\AgdaGeneralizable{𝒞}\AgdaSpace{}%
\AgdaGeneralizable{𝒟}\AgdaSpace{}%
\AgdaGeneralizable{𝒱}\AgdaSpace{}%
\AgdaGeneralizable{𝒲}\AgdaSpace{}%
\AgdaSymbol{:}\AgdaSpace{}%
\AgdaFunction{Scope}\<%
\\
\>[8]\AgdaGeneralizable{v}\AgdaSpace{}%
\AgdaSymbol{:}\AgdaSpace{}%
\AgdaFunction{𝓥}\AgdaSpace{}%
\AgdaGeneralizable{Γ}\AgdaSpace{}%
\AgdaGeneralizable{i}\<%
\end{code}

From this section on, we move from \emph{doing} things to \emph{proving} things.
For instance, in a classic proof of the Church-Rosser theorem, a technique
is used that ``colors'' specific $\lambda$'s to track their behavior in
reduction. An indispensable lemma of that proof is that
\emph{substitution commutes with color erasure}. Suppose $\Lambda$
is the set of uncolored terms, and $\overline\Lambda$ is the set of
terms with color. $\lfloor \cdot \rfloor : \overline\Lambda \to \Lambda$
erases the colors. Then the lemma to be proved is
$$ \left\lfloor t (x \mapsto s) \right\rfloor = \lfloor t \rfloor ( x \mapsto \lfloor s \rfloor ). $$
This looks suspiciously similar to \emph{homomorphisms} in abstract algebra.
For example a group homomorphism $\phi$ satisfies
$$\phi(x \cdot y) = \phi(x) \cdot \phi(y).$$
One crucial difference: The definition of group homomorphisms is directly
concerned about group multiplication, which is a ``primitive'' concept; On
the other hand, substitution
is defined in terms of \AgdaField{var} and \AgdaField{map}.
So it is clear that we need to come up with an equation in terms of these
two primitives.

\AgdaTarget{Hom}
\begin{code}%
\>[4]\AgdaKeyword{record}\AgdaSpace{}%
\AgdaRecord{Hom}\AgdaSpace{}%
\AgdaSymbol{⦃}\AgdaSpace{}%
\AgdaBound{𝒞ˢ}\AgdaSpace{}%
\AgdaSymbol{:}\AgdaSpace{}%
\AgdaRecord{Syntax}\AgdaSpace{}%
\AgdaGeneralizable{𝒞}\AgdaSpace{}%
\AgdaSymbol{⦄}\AgdaSpace{}%
\AgdaSymbol{⦃}\AgdaSpace{}%
\AgdaBound{𝒟ˢ}\AgdaSpace{}%
\AgdaSymbol{:}\AgdaSpace{}%
\AgdaRecord{Syntax}\AgdaSpace{}%
\AgdaGeneralizable{𝒟}\AgdaSpace{}%
\AgdaSymbol{⦄}\<%
\\
\>[4][@{}l@{\AgdaIndent{0}}]%
\>[8]\AgdaSymbol{(}\AgdaBound{𝑓}\AgdaSpace{}%
\AgdaSymbol{:}\AgdaSpace{}%
\AgdaOperator{\AgdaFunction{[}}\AgdaSpace{}%
\AgdaGeneralizable{𝒞}\AgdaSpace{}%
\AgdaOperator{\AgdaFunction{⇒}}\AgdaSpace{}%
\AgdaGeneralizable{𝒟}\AgdaSpace{}%
\AgdaOperator{\AgdaFunction{]}}\AgdaSymbol{)}\AgdaSpace{}%
\AgdaSymbol{:}\AgdaSpace{}%
\AgdaPrimitive{Set₁}\AgdaSpace{}%
\AgdaKeyword{where}\<%
\\
\>[8]\AgdaKeyword{private}\AgdaSpace{}%
\AgdaKeyword{instance}\<%
\\
\>[8][@{}l@{\AgdaIndent{0}}]%
\>[12]\AgdaFunction{\AgdaUnderscore{}}\AgdaSpace{}%
\AgdaSymbol{=}\AgdaSpace{}%
\AgdaFunction{Syntaxʷ}\AgdaSpace{}%
\AgdaSymbol{⦃}\AgdaSpace{}%
\AgdaBound{𝒞ˢ}\AgdaSpace{}%
\AgdaSymbol{⦄}\<%
\\
\>[12]\AgdaFunction{\AgdaUnderscore{}}\AgdaSpace{}%
\AgdaSymbol{=}\AgdaSpace{}%
\AgdaFunction{Syntaxʷ}\AgdaSpace{}%
\AgdaSymbol{⦃}\AgdaSpace{}%
\AgdaBound{𝒟ˢ}\AgdaSpace{}%
\AgdaSymbol{⦄}\<%
\\
\>[8]\AgdaKeyword{field}\<%
\end{code}

What may we say about \AgdaArgument{𝑓}? For \AgdaField{var} it's easy.
We just need to take care of some implicit arguments:
\AgdaTarget{Hvar}
\begin{code}%
\>[8][@{}l@{\AgdaIndent{1}}]%
\>[12]\AgdaField{Hvar}\AgdaSpace{}%
\AgdaSymbol{:}\AgdaSpace{}%
\AgdaBound{𝑓}\AgdaSpace{}%
\AgdaSymbol{\{}\AgdaGeneralizable{Γ}\AgdaSymbol{\}}\AgdaSpace{}%
\AgdaSymbol{\{}\AgdaGeneralizable{i}\AgdaSymbol{\}}\AgdaSpace{}%
\AgdaSymbol{(}\AgdaField{var}\AgdaSpace{}%
\AgdaGeneralizable{v}\AgdaSymbol{)}\AgdaSpace{}%
\AgdaOperator{\AgdaDatatype{≡}}\AgdaSpace{}%
\AgdaField{var}\AgdaSpace{}%
\AgdaGeneralizable{v}\<%
\end{code}

But there is some difficulty for \AgdaField{map}, because it is a
\emph{higher-order} function, acting on other functions. How do we write
out the homomorphism requirements for high-order structures? For concreteness,
let's take an example.

We define a \textbf{fixoid} to be a set $X$ together with a functional
$F : (X \to X) \to X$ on it. This has no practical use whatsoever,%
\footnote{Actually, if we impose that $F$ sends functions to their
fixpoints: $$F(f) = f(F(f)),$$then this becomes what is called a
\textbf{fixed-point space}. It is trivial for discrete spaces, but once
we impose a topology it becomes interesting. For example, any closed interval
endowed with the real topology is a fixed-point space.}
and is only
intended as an example for us to discuss what higher-order homomorphisms
should be like.

Take two fixoids, $(X, F)$ and $(Y, G)$. What makes a function $\phi : X \to Y$
a homomorphism? The first reasonable guess is that we have an equation
of the form
$$ \phi(F(?)) = G(?). $$
What should we put in the arguments? There should be an $f : X \to X$
and a $g : Y \to Y$. But surely they can't be arbitrary. What should be
the relation between them? Looking at the type signatures, let's go for
the obvious:
$$\forall x, \quad\phi(f(x)) = g(\phi(x)).$$
So, if we collect everything, we get the definition:

A \textbf{homomorphism} between fixoids $(X, F)$ and $(Y, G)$ is a function
$\phi : X \to Y$ such that for every pair of functions $f : X \to X$
and $g : Y \to Y$,
$$\phi \circ f = g\circ\phi  \implies  \phi(F(f)) = G(g).$$

In fact, any other ``reasonable'' conditions we may impose are already
implied by this one! The proof is quite combinatorial, and is left as an
exercise for the reader.

\section{Incoherent Coherences}

In a similar spirit, after meditating at the type signature for a while,
one may come up with such a condition for \AgdaField{map}:
\begin{code}%
\>[12]\AgdaField{Hnat}\AgdaSpace{}%
\AgdaSymbol{:}\AgdaSpace{}%
\AgdaSymbol{⦃}\AgdaSpace{}%
\AgdaBound{𝒜ʷ}\AgdaSpace{}%
\AgdaSymbol{:}\AgdaSpace{}%
\AgdaRecord{Weakening}\AgdaSpace{}%
\AgdaGeneralizable{𝒜}\AgdaSpace{}%
\AgdaSymbol{⦄}\AgdaSpace{}%
\AgdaSymbol{(}\AgdaBound{δ}\AgdaSpace{}%
\AgdaSymbol{:}\AgdaSpace{}%
\AgdaOperator{\AgdaFunction{[}}\AgdaSpace{}%
\AgdaGeneralizable{𝒜}\AgdaSpace{}%
\AgdaOperator{\AgdaFunction{⇒}}\AgdaSpace{}%
\AgdaBound{𝒞}\AgdaSpace{}%
\AgdaOperator{\AgdaFunction{]}}\AgdaSymbol{)}\<%
\\
\>[12][@{}l@{\AgdaIndent{0}}]%
\>[16]\AgdaSymbol{→}\AgdaSpace{}%
\AgdaSymbol{∀}\AgdaSpace{}%
\AgdaSymbol{\{}\AgdaBound{Γ}\AgdaSpace{}%
\AgdaBound{Δ}\AgdaSymbol{\}}\AgdaSpace{}%
\AgdaSymbol{(}\AgdaBound{σ}\AgdaSpace{}%
\AgdaSymbol{:}\AgdaSpace{}%
\AgdaSymbol{(}\AgdaFunction{𝓥}\AgdaSpace{}%
\AgdaOperator{\AgdaFunction{⇒}}\AgdaSpace{}%
\AgdaGeneralizable{𝒜}\AgdaSymbol{)}\AgdaSpace{}%
\AgdaBound{Γ}\AgdaSpace{}%
\AgdaBound{Δ}\AgdaSymbol{)}\AgdaSpace{}%
\AgdaSymbol{\{}\AgdaBound{i}\AgdaSymbol{\}}\AgdaSpace{}%
\AgdaSymbol{(}\AgdaBound{v}\AgdaSpace{}%
\AgdaSymbol{:}\AgdaSpace{}%
\AgdaBound{𝒞}\AgdaSpace{}%
\AgdaBound{Γ}\AgdaSpace{}%
\AgdaBound{i}\AgdaSymbol{)}\<%
\\
\>[16]\AgdaSymbol{→}\AgdaSpace{}%
\AgdaBound{𝑓}\AgdaSpace{}%
\AgdaSymbol{(}\AgdaField{map}\AgdaSpace{}%
\AgdaBound{δ}\AgdaSpace{}%
\AgdaBound{σ}\AgdaSpace{}%
\AgdaBound{v}\AgdaSymbol{)}\AgdaSpace{}%
\AgdaOperator{\AgdaDatatype{≡}}\AgdaSpace{}%
\AgdaSymbol{(}\AgdaField{map}\AgdaSpace{}%
\AgdaSymbol{(}\AgdaBound{𝑓}\AgdaSpace{}%
\AgdaOperator{\AgdaFunction{∘}}\AgdaSpace{}%
\AgdaBound{δ}\AgdaSymbol{)}\AgdaSpace{}%
\AgdaBound{σ}\AgdaSymbol{)}\AgdaSpace{}%
\AgdaSymbol{(}\AgdaBound{𝑓}\AgdaSpace{}%
\AgdaBound{v}\AgdaSymbol{)}\<%
\end{code}

It looks quite intimidating, but the first two lines are just setup.
It is pretty much what is expected, but unfortunately it is not enough.
In the degenerate case that $𝒜 = 𝒞$, it turns out that we need an additional
rule...

\begin{code}%
\>[12]\AgdaField{Hpol}\AgdaSpace{}%
\AgdaSymbol{:}\AgdaSpace{}%
\AgdaSymbol{(}\AgdaBound{δ}\AgdaSpace{}%
\AgdaSymbol{:}\AgdaSpace{}%
\AgdaOperator{\AgdaFunction{[}}\AgdaSpace{}%
\AgdaBound{𝒞}\AgdaSpace{}%
\AgdaOperator{\AgdaFunction{⇒}}\AgdaSpace{}%
\AgdaBound{𝒞}\AgdaSpace{}%
\AgdaOperator{\AgdaFunction{]}}\AgdaSymbol{)}\AgdaSpace{}%
\AgdaSymbol{(}\AgdaBound{δ'}\AgdaSpace{}%
\AgdaSymbol{:}\AgdaSpace{}%
\AgdaOperator{\AgdaFunction{[}}\AgdaSpace{}%
\AgdaBound{𝒟}\AgdaSpace{}%
\AgdaOperator{\AgdaFunction{⇒}}\AgdaSpace{}%
\AgdaBound{𝒟}\AgdaSpace{}%
\AgdaOperator{\AgdaFunction{]}}\AgdaSymbol{)}\<%
\\
\>[12][@{}l@{\AgdaIndent{0}}]%
\>[16]\AgdaSymbol{→}%
\>[1169I]\AgdaSymbol{(}\AgdaBound{nat}\AgdaSpace{}%
\AgdaSymbol{:}\AgdaSpace{}%
\AgdaSymbol{∀}\AgdaSpace{}%
\AgdaSymbol{\{}\AgdaBound{Γ}\AgdaSpace{}%
\AgdaBound{σ}\AgdaSymbol{\}}\AgdaSpace{}%
\AgdaSymbol{(}\AgdaBound{t}\AgdaSpace{}%
\AgdaSymbol{:}\AgdaSpace{}%
\AgdaBound{𝒞}\AgdaSpace{}%
\AgdaBound{Γ}\AgdaSpace{}%
\AgdaBound{σ}\AgdaSymbol{)}\<%
\\
\>[1169I][@{}l@{\AgdaIndent{0}}]%
\>[20]\AgdaSymbol{→}\AgdaSpace{}%
\AgdaBound{𝑓}\AgdaSpace{}%
\AgdaSymbol{(}\AgdaBound{δ}\AgdaSpace{}%
\AgdaBound{t}\AgdaSymbol{)}\AgdaSpace{}%
\AgdaOperator{\AgdaDatatype{≡}}\AgdaSpace{}%
\AgdaBound{δ'}\AgdaSpace{}%
\AgdaSymbol{(}\AgdaBound{𝑓}\AgdaSpace{}%
\AgdaBound{t}\AgdaSymbol{))}%
\>[44]\AgdaComment{----\ !}\<%
\\
\>[16]\AgdaSymbol{→}%
\>[1186I]\AgdaSymbol{(}\AgdaBound{wk}\AgdaSpace{}%
\AgdaSymbol{:}\AgdaSpace{}%
\AgdaSymbol{∀}\AgdaSpace{}%
\AgdaSymbol{\{}\AgdaBound{Γ}\AgdaSpace{}%
\AgdaBound{Δ}\AgdaSymbol{\}}\AgdaSpace{}%
\AgdaSymbol{(}\AgdaBound{σ}\AgdaSpace{}%
\AgdaSymbol{:}\AgdaSpace{}%
\AgdaSymbol{(}\AgdaFunction{𝓥}\AgdaSpace{}%
\AgdaOperator{\AgdaFunction{⇒}}\AgdaSpace{}%
\AgdaBound{𝒞}\AgdaSymbol{)}\AgdaSpace{}%
\AgdaBound{Γ}\AgdaSpace{}%
\AgdaBound{Δ}\AgdaSymbol{)}\AgdaSpace{}%
\AgdaSymbol{\{}\AgdaBound{i}\AgdaSpace{}%
\AgdaBound{j}\AgdaSymbol{\}}\AgdaSpace{}%
\AgdaSymbol{(}\AgdaBound{v}\AgdaSpace{}%
\AgdaSymbol{:}\AgdaSpace{}%
\AgdaFunction{𝓥}\AgdaSpace{}%
\AgdaSymbol{\AgdaUnderscore{}}\AgdaSpace{}%
\AgdaBound{i}\AgdaSymbol{)}\<%
\\
\>[1186I][@{}l@{\AgdaIndent{0}}]%
\>[20]\AgdaSymbol{→}\AgdaSpace{}%
\AgdaBound{𝑓}\AgdaSpace{}%
\AgdaSymbol{((}\AgdaBound{σ}\AgdaSpace{}%
\AgdaOperator{\AgdaFunction{≪}}\AgdaSpace{}%
\AgdaBound{j}\AgdaSymbol{)}\AgdaSpace{}%
\AgdaBound{v}\AgdaSymbol{)}\AgdaSpace{}%
\AgdaOperator{\AgdaDatatype{≡}}\AgdaSpace{}%
\AgdaSymbol{((}\AgdaBound{𝑓}\AgdaSpace{}%
\AgdaOperator{\AgdaFunction{∘}}\AgdaSpace{}%
\AgdaBound{σ}\AgdaSymbol{)}\AgdaSpace{}%
\AgdaOperator{\AgdaFunction{≪}}\AgdaSpace{}%
\AgdaBound{j}\AgdaSymbol{)}\AgdaSpace{}%
\AgdaBound{v}\AgdaSymbol{)}\<%
\\
\>[16]\AgdaSymbol{→}\AgdaSpace{}%
\AgdaSymbol{∀}\AgdaSpace{}%
\AgdaSymbol{\{}\AgdaBound{Γ}\AgdaSpace{}%
\AgdaBound{Δ}\AgdaSymbol{\}}\AgdaSpace{}%
\AgdaSymbol{(}\AgdaBound{σ}\AgdaSpace{}%
\AgdaSymbol{:}\AgdaSpace{}%
\AgdaSymbol{(}\AgdaFunction{𝓥}\AgdaSpace{}%
\AgdaOperator{\AgdaFunction{⇒}}\AgdaSpace{}%
\AgdaBound{𝒞}\AgdaSymbol{)}\AgdaSpace{}%
\AgdaBound{Γ}\AgdaSpace{}%
\AgdaBound{Δ}\AgdaSymbol{)}\AgdaSpace{}%
\AgdaSymbol{\{}\AgdaBound{i}\AgdaSymbol{\}}\AgdaSpace{}%
\AgdaSymbol{(}\AgdaBound{t}\AgdaSpace{}%
\AgdaSymbol{:}\AgdaSpace{}%
\AgdaBound{𝒞}\AgdaSpace{}%
\AgdaBound{Γ}\AgdaSpace{}%
\AgdaBound{i}\AgdaSymbol{)}\<%
\\
\>[16]\AgdaSymbol{→}\AgdaSpace{}%
\AgdaBound{𝑓}\AgdaSpace{}%
\AgdaSymbol{(}\AgdaField{map}\AgdaSpace{}%
\AgdaBound{δ}\AgdaSpace{}%
\AgdaBound{σ}\AgdaSpace{}%
\AgdaBound{t}\AgdaSymbol{)}\AgdaSpace{}%
\AgdaOperator{\AgdaDatatype{≡}}\AgdaSpace{}%
\AgdaField{map}\AgdaSpace{}%
\AgdaBound{δ'}\AgdaSpace{}%
\AgdaSymbol{(}\AgdaBound{𝑓}\AgdaSpace{}%
\AgdaOperator{\AgdaFunction{∘}}\AgdaSpace{}%
\AgdaBound{σ}\AgdaSymbol{)}\AgdaSpace{}%
\AgdaSymbol{(}\AgdaBound{𝑓}\AgdaSpace{}%
\AgdaBound{t}\AgdaSymbol{)}%
\>[57]\AgdaComment{----\ !}\<%
\end{code}

... It's even more intimidating. But only the two lines marked by comments
matters. The first says that the two \AgdaArgument{δ}'s satisfy a similar
condition to the equation for $f$ and $g$ in the case of fixoids.
The second is \emph{slightly} different to \AgdaField{Hnat}, but also reasonable.

The argument \AgdaArgument{wk} is actually not necessary, and we will
eliminate it. But when implementing this typeclass method, it helps to
make inductions go through.

\begin{code}[hide]%
\>[8]\AgdaKeyword{private}\<%
\\
\>[8][@{}l@{\AgdaIndent{0}}]%
\>[12]\AgdaOperator{\AgdaFunction{fH𝕫/\AgdaUnderscore{}}}\AgdaSpace{}%
\AgdaSymbol{:}\AgdaSpace{}%
\AgdaSymbol{(}\AgdaBound{t}\AgdaSpace{}%
\AgdaSymbol{:}\AgdaSpace{}%
\AgdaBound{𝒞}\AgdaSpace{}%
\AgdaGeneralizable{Γ}\AgdaSpace{}%
\AgdaGeneralizable{i}\AgdaSymbol{)}\AgdaSpace{}%
\AgdaSymbol{(}\AgdaBound{v}\AgdaSpace{}%
\AgdaSymbol{:}\AgdaSpace{}%
\AgdaFunction{𝓥}\AgdaSpace{}%
\AgdaSymbol{(}\AgdaGeneralizable{Γ}\AgdaSpace{}%
\AgdaOperator{\AgdaInductiveConstructor{◂}}\AgdaSpace{}%
\AgdaGeneralizable{i}\AgdaSymbol{)}\AgdaSpace{}%
\AgdaGeneralizable{j}\AgdaSymbol{)}\<%
\\
\>[12][@{}l@{\AgdaIndent{0}}]%
\>[16]\AgdaSymbol{→}\AgdaSpace{}%
\AgdaBound{𝑓}\AgdaSpace{}%
\AgdaSymbol{((}\AgdaOperator{\AgdaFunction{𝕫/}}\AgdaSpace{}%
\AgdaBound{t}\AgdaSymbol{)}\AgdaSpace{}%
\AgdaBound{v}\AgdaSymbol{)}\AgdaSpace{}%
\AgdaOperator{\AgdaDatatype{≡}}\AgdaSpace{}%
\AgdaSymbol{(}\AgdaOperator{\AgdaFunction{𝕫/}}\AgdaSpace{}%
\AgdaBound{𝑓}\AgdaSpace{}%
\AgdaBound{t}\AgdaSymbol{)}\AgdaSpace{}%
\AgdaBound{v}\<%
\\
\>[12]\AgdaSymbol{(}\AgdaOperator{\AgdaFunction{fH𝕫/}}\AgdaSpace{}%
\AgdaBound{t}\AgdaSymbol{)}\AgdaSpace{}%
\AgdaInductiveConstructor{𝕫}\AgdaSpace{}%
\AgdaSymbol{=}\AgdaSpace{}%
\AgdaInductiveConstructor{refl}\<%
\\
\>[12]\AgdaSymbol{(}\AgdaOperator{\AgdaFunction{fH𝕫/}}\AgdaSpace{}%
\AgdaBound{t}\AgdaSymbol{)}\AgdaSpace{}%
\AgdaSymbol{(}\AgdaOperator{\AgdaInductiveConstructor{𝕤}}\AgdaSpace{}%
\AgdaBound{v}\AgdaSymbol{)}\AgdaSpace{}%
\AgdaSymbol{=}\AgdaSpace{}%
\AgdaField{Hvar}\<%
\\
\\[\AgdaEmptyExtraSkip]%
\>[8]\AgdaOperator{\AgdaFunction{H𝕫/\AgdaUnderscore{}}}\AgdaSpace{}%
\AgdaSymbol{:}\AgdaSpace{}%
\AgdaSymbol{∀}\AgdaSpace{}%
\AgdaSymbol{(}\AgdaBound{t}\AgdaSpace{}%
\AgdaSymbol{:}\AgdaSpace{}%
\AgdaBound{𝒞}\AgdaSpace{}%
\AgdaGeneralizable{Γ}\AgdaSpace{}%
\AgdaGeneralizable{i}\AgdaSymbol{)}\<%
\\
\>[8][@{}l@{\AgdaIndent{0}}]%
\>[12]\AgdaSymbol{→}\AgdaSpace{}%
\AgdaSymbol{(λ}\AgdaSpace{}%
\AgdaSymbol{\{}\AgdaBound{j}\AgdaSymbol{\}}\AgdaSpace{}%
\AgdaSymbol{→}\AgdaSpace{}%
\AgdaBound{𝑓}\AgdaSpace{}%
\AgdaSymbol{\{}\AgdaArgument{i}\AgdaSpace{}%
\AgdaSymbol{=}\AgdaSpace{}%
\AgdaBound{j}\AgdaSymbol{\}}\AgdaSpace{}%
\AgdaOperator{\AgdaFunction{∘}}\AgdaSpace{}%
\AgdaSymbol{(}\AgdaOperator{\AgdaFunction{𝕫/}}\AgdaSpace{}%
\AgdaBound{t}\AgdaSymbol{))}\AgdaSpace{}%
\AgdaOperator{\AgdaDatatype{≡}}\AgdaSpace{}%
\AgdaOperator{\AgdaFunction{𝕫/}}\AgdaSpace{}%
\AgdaBound{𝑓}\AgdaSpace{}%
\AgdaBound{t}\<%
\\
\>[8]\AgdaOperator{\AgdaFunction{H𝕫/}}\AgdaSpace{}%
\AgdaBound{t}\AgdaSpace{}%
\AgdaSymbol{=}\AgdaSpace{}%
\AgdaPostulate{funext'}\AgdaSpace{}%
\AgdaSymbol{\textbackslash{}}\AgdaSpace{}%
\AgdaBound{\AgdaUnderscore{}}\AgdaSpace{}%
\AgdaSymbol{→}\AgdaSpace{}%
\AgdaPostulate{funext}\AgdaSpace{}%
\AgdaSymbol{(}\AgdaOperator{\AgdaFunction{fH𝕫/}}\AgdaSpace{}%
\AgdaBound{t}\AgdaSymbol{)}\<%
\\
\\[\AgdaEmptyExtraSkip]%
\>[8]\AgdaKeyword{private}\<%
\\
\>[8][@{}l@{\AgdaIndent{0}}]%
\>[12]\AgdaFunction{fHvar}\AgdaSpace{}%
\AgdaSymbol{:}\AgdaSpace{}%
\AgdaSymbol{(\textbackslash{}\{}\AgdaBound{Γ}\AgdaSpace{}%
\AgdaBound{i}\AgdaSymbol{\}}\AgdaSpace{}%
\AgdaSymbol{→}\AgdaSpace{}%
\AgdaBound{𝑓}\AgdaSpace{}%
\AgdaSymbol{\{}\AgdaBound{Γ}\AgdaSymbol{\}}\AgdaSpace{}%
\AgdaSymbol{\{}\AgdaBound{i}\AgdaSymbol{\}}\AgdaSpace{}%
\AgdaOperator{\AgdaFunction{∘}}\AgdaSpace{}%
\AgdaField{var}\AgdaSymbol{)}\AgdaSpace{}%
\AgdaOperator{\AgdaDatatype{≡}}\AgdaSpace{}%
\AgdaField{var}\<%
\\
\>[12]\AgdaFunction{fHvar}\AgdaSpace{}%
\AgdaSymbol{=}\<%
\\
\>[12][@{}l@{\AgdaIndent{0}}]%
\>[16]\AgdaPostulate{funext'}\AgdaSpace{}%
\AgdaSymbol{\textbackslash{}}\AgdaSpace{}%
\AgdaBound{\AgdaUnderscore{}}\AgdaSpace{}%
\AgdaSymbol{→}\<%
\\
\>[16]\AgdaPostulate{funext'}\AgdaSpace{}%
\AgdaSymbol{\textbackslash{}}\AgdaSpace{}%
\AgdaBound{\AgdaUnderscore{}}\AgdaSpace{}%
\AgdaSymbol{→}\<%
\\
\>[16]\AgdaPostulate{funext}%
\>[24]\AgdaSymbol{\textbackslash{}}\AgdaSpace{}%
\AgdaBound{\AgdaUnderscore{}}\AgdaSpace{}%
\AgdaSymbol{→}\AgdaSpace{}%
\AgdaField{Hvar}\<%
\end{code}

But after some straightforward definitions, we can prove the desired lemma
for \AgdaFunction{rename}:
\begin{code}%
\>[8]\AgdaFunction{Hrename}\AgdaSpace{}%
\AgdaSymbol{:}\AgdaSpace{}%
\AgdaSymbol{∀}\AgdaSpace{}%
\AgdaSymbol{\{}\AgdaBound{Γ}\AgdaSpace{}%
\AgdaBound{Δ}\AgdaSymbol{\}}\AgdaSpace{}%
\AgdaSymbol{(}\AgdaBound{ρ}\AgdaSpace{}%
\AgdaSymbol{:}\AgdaSpace{}%
\AgdaSymbol{(}\AgdaFunction{𝓥}\AgdaSpace{}%
\AgdaOperator{\AgdaFunction{⇒}}\AgdaSpace{}%
\AgdaFunction{𝓥}\AgdaSymbol{)}\AgdaSpace{}%
\AgdaBound{Γ}\AgdaSpace{}%
\AgdaBound{Δ}\AgdaSymbol{)}\<%
\\
\>[8][@{}l@{\AgdaIndent{0}}]%
\>[12]\AgdaSymbol{→}\AgdaSpace{}%
\AgdaSymbol{∀}\AgdaSpace{}%
\AgdaSymbol{\{}\AgdaBound{i}\AgdaSymbol{\}}\AgdaSpace{}%
\AgdaSymbol{(}\AgdaBound{t}\AgdaSpace{}%
\AgdaSymbol{:}\AgdaSpace{}%
\AgdaBound{𝒞}\AgdaSpace{}%
\AgdaBound{Γ}\AgdaSpace{}%
\AgdaBound{i}\AgdaSymbol{)}\<%
\\
\>[12]\AgdaSymbol{→}\AgdaSpace{}%
\AgdaBound{𝑓}\AgdaSpace{}%
\AgdaSymbol{(}\AgdaFunction{rename}\AgdaSpace{}%
\AgdaBound{ρ}\AgdaSpace{}%
\AgdaBound{t}\AgdaSymbol{)}\AgdaSpace{}%
\AgdaOperator{\AgdaDatatype{≡}}\AgdaSpace{}%
\AgdaFunction{rename}\AgdaSpace{}%
\AgdaBound{ρ}\AgdaSpace{}%
\AgdaSymbol{(}\AgdaBound{𝑓}\AgdaSpace{}%
\AgdaBound{t}\AgdaSymbol{)}\<%
\\
\>[8]\AgdaFunction{Hrename}\AgdaSpace{}%
\AgdaBound{ρ}\AgdaSpace{}%
\AgdaBound{t}\AgdaSpace{}%
\AgdaKeyword{rewrite}\AgdaSpace{}%
\AgdaFunction{symm}\AgdaSpace{}%
\AgdaFunction{fHvar}\AgdaSpace{}%
\AgdaSymbol{=}\AgdaSpace{}%
\AgdaField{Hnat}\AgdaSpace{}%
\AgdaField{var}\AgdaSpace{}%
\AgdaBound{ρ}\AgdaSpace{}%
\AgdaBound{t}\<%
\end{code}
\begin{code}[hide]%
\>[8]\AgdaFunction{Hweaken}\AgdaSpace{}%
\AgdaSymbol{:}\AgdaSpace{}%
\AgdaSymbol{∀}\AgdaSpace{}%
\AgdaSymbol{(}\AgdaBound{σ}\AgdaSpace{}%
\AgdaSymbol{:}\AgdaSpace{}%
\AgdaSymbol{(}\AgdaFunction{𝓥}\AgdaSpace{}%
\AgdaOperator{\AgdaFunction{⇒}}\AgdaSpace{}%
\AgdaBound{𝒞}\AgdaSymbol{)}\AgdaSpace{}%
\AgdaGeneralizable{Γ}\AgdaSpace{}%
\AgdaGeneralizable{Δ}\AgdaSymbol{)}\AgdaSpace{}%
\AgdaSymbol{\{}\AgdaBound{i}\AgdaSpace{}%
\AgdaBound{j}\AgdaSymbol{\}}\AgdaSpace{}%
\AgdaSymbol{(}\AgdaBound{v}\AgdaSpace{}%
\AgdaSymbol{:}\AgdaSpace{}%
\AgdaGeneralizable{Γ}\AgdaSpace{}%
\AgdaOperator{\AgdaInductiveConstructor{◂}}\AgdaSpace{}%
\AgdaBound{i}\AgdaSpace{}%
\AgdaOperator{\AgdaDatatype{∋}}\AgdaSpace{}%
\AgdaBound{j}\AgdaSymbol{)}\<%
\\
\>[8][@{}l@{\AgdaIndent{0}}]%
\>[12]\AgdaSymbol{→}\AgdaSpace{}%
\AgdaBound{𝑓}\AgdaSpace{}%
\AgdaSymbol{((}\AgdaBound{σ}\AgdaSpace{}%
\AgdaOperator{\AgdaFunction{≪}}\AgdaSpace{}%
\AgdaBound{i}\AgdaSymbol{)}\AgdaSpace{}%
\AgdaBound{v}\AgdaSymbol{)}\AgdaSpace{}%
\AgdaOperator{\AgdaDatatype{≡}}\AgdaSpace{}%
\AgdaSymbol{((}\AgdaBound{𝑓}\AgdaSpace{}%
\AgdaOperator{\AgdaFunction{∘}}\AgdaSpace{}%
\AgdaBound{σ}\AgdaSymbol{)}\AgdaSpace{}%
\AgdaOperator{\AgdaFunction{≪}}\AgdaSpace{}%
\AgdaBound{i}\AgdaSymbol{)}\AgdaSpace{}%
\AgdaBound{v}\<%
\\
\>[8]\AgdaFunction{Hweaken}\AgdaSpace{}%
\AgdaBound{σ}\AgdaSpace{}%
\AgdaInductiveConstructor{𝕫}\AgdaSpace{}%
\AgdaSymbol{=}\AgdaSpace{}%
\AgdaField{Hvar}\<%
\\
\>[8]\AgdaFunction{Hweaken}\AgdaSpace{}%
\AgdaBound{σ}\AgdaSpace{}%
\AgdaSymbol{(}\AgdaOperator{\AgdaInductiveConstructor{𝕤}}\AgdaSpace{}%
\AgdaBound{v}\AgdaSymbol{)}\AgdaSpace{}%
\AgdaSymbol{=}\AgdaSpace{}%
\AgdaFunction{Hrename}\AgdaSpace{}%
\AgdaOperator{\AgdaInductiveConstructor{𝕤\AgdaUnderscore{}}}\AgdaSpace{}%
\AgdaSymbol{(}\AgdaBound{σ}\AgdaSpace{}%
\AgdaBound{v}\AgdaSymbol{)}\<%
\\
\\[\AgdaEmptyExtraSkip]%
\>[8]\AgdaKeyword{private}\<%
\\
\>[8][@{}l@{\AgdaIndent{0}}]%
\>[12]\AgdaFunction{fHweaken}\AgdaSpace{}%
\AgdaSymbol{:}\AgdaSpace{}%
\AgdaSymbol{(}\AgdaBound{σ}\AgdaSpace{}%
\AgdaSymbol{:}\AgdaSpace{}%
\AgdaSymbol{(}\AgdaFunction{𝓥}\AgdaSpace{}%
\AgdaOperator{\AgdaFunction{⇒}}\AgdaSpace{}%
\AgdaBound{𝒞}\AgdaSymbol{)}\AgdaSpace{}%
\AgdaGeneralizable{Γ}\AgdaSpace{}%
\AgdaGeneralizable{Δ}\AgdaSymbol{)}\<%
\\
\>[12][@{}l@{\AgdaIndent{0}}]%
\>[16]\AgdaSymbol{→}\AgdaSpace{}%
\AgdaBound{𝑓}\AgdaSpace{}%
\AgdaSymbol{\{}\AgdaArgument{i}\AgdaSpace{}%
\AgdaSymbol{=}\AgdaSpace{}%
\AgdaGeneralizable{j}\AgdaSymbol{\}}\AgdaSpace{}%
\AgdaOperator{\AgdaFunction{∘}}\AgdaSpace{}%
\AgdaSymbol{(}\AgdaBound{σ}\AgdaSpace{}%
\AgdaOperator{\AgdaFunction{≪}}\AgdaSpace{}%
\AgdaGeneralizable{i}\AgdaSymbol{)}\AgdaSpace{}%
\AgdaOperator{\AgdaDatatype{≡}}\AgdaSpace{}%
\AgdaSymbol{(}\AgdaBound{𝑓}\AgdaSpace{}%
\AgdaOperator{\AgdaFunction{∘}}\AgdaSpace{}%
\AgdaBound{σ}\AgdaSymbol{)}\AgdaSpace{}%
\AgdaOperator{\AgdaFunction{≪}}\AgdaSpace{}%
\AgdaGeneralizable{i}\<%
\\
\>[12]\AgdaFunction{fHweaken}\AgdaSpace{}%
\AgdaBound{σ}\AgdaSpace{}%
\AgdaSymbol{=}\AgdaSpace{}%
\AgdaPostulate{funext}\AgdaSpace{}%
\AgdaSymbol{(}\AgdaFunction{Hweaken}\AgdaSpace{}%
\AgdaBound{σ}\AgdaSymbol{)}\<%
\\
\\[\AgdaEmptyExtraSkip]%
\>[8]\AgdaFunction{Hmap}\AgdaSpace{}%
\AgdaSymbol{:}\AgdaSpace{}%
\AgdaSymbol{(}\AgdaBound{δ}\AgdaSpace{}%
\AgdaSymbol{:}\AgdaSpace{}%
\AgdaOperator{\AgdaFunction{[}}\AgdaSpace{}%
\AgdaBound{𝒞}\AgdaSpace{}%
\AgdaOperator{\AgdaFunction{⇒}}\AgdaSpace{}%
\AgdaBound{𝒞}\AgdaSpace{}%
\AgdaOperator{\AgdaFunction{]}}\AgdaSymbol{)}\AgdaSpace{}%
\AgdaSymbol{(}\AgdaBound{δ'}\AgdaSpace{}%
\AgdaSymbol{:}\AgdaSpace{}%
\AgdaOperator{\AgdaFunction{[}}\AgdaSpace{}%
\AgdaBound{𝒟}\AgdaSpace{}%
\AgdaOperator{\AgdaFunction{⇒}}\AgdaSpace{}%
\AgdaBound{𝒟}\AgdaSpace{}%
\AgdaOperator{\AgdaFunction{]}}\AgdaSymbol{)}\<%
\\
\>[8][@{}l@{\AgdaIndent{0}}]%
\>[12]\AgdaSymbol{→}%
\>[1449I]\AgdaSymbol{(}\AgdaBound{nat}\AgdaSpace{}%
\AgdaSymbol{:}\AgdaSpace{}%
\AgdaSymbol{∀}\AgdaSpace{}%
\AgdaSymbol{\{}\AgdaBound{Γ}\AgdaSpace{}%
\AgdaBound{σ}\AgdaSymbol{\}}\AgdaSpace{}%
\AgdaSymbol{(}\AgdaBound{t}\AgdaSpace{}%
\AgdaSymbol{:}\AgdaSpace{}%
\AgdaBound{𝒞}\AgdaSpace{}%
\AgdaBound{Γ}\AgdaSpace{}%
\AgdaBound{σ}\AgdaSymbol{)}\<%
\\
\>[1449I][@{}l@{\AgdaIndent{0}}]%
\>[16]\AgdaSymbol{→}\AgdaSpace{}%
\AgdaBound{𝑓}\AgdaSpace{}%
\AgdaSymbol{(}\AgdaBound{δ}\AgdaSpace{}%
\AgdaBound{t}\AgdaSymbol{)}\AgdaSpace{}%
\AgdaOperator{\AgdaDatatype{≡}}\AgdaSpace{}%
\AgdaBound{δ'}\AgdaSpace{}%
\AgdaSymbol{(}\AgdaBound{𝑓}\AgdaSpace{}%
\AgdaBound{t}\AgdaSymbol{))}\<%
\\
\>[12]\AgdaSymbol{→}\AgdaSpace{}%
\AgdaSymbol{∀}\AgdaSpace{}%
\AgdaSymbol{\{}\AgdaBound{Γ}\AgdaSpace{}%
\AgdaBound{Δ}\AgdaSymbol{\}}\AgdaSpace{}%
\AgdaSymbol{(}\AgdaBound{σ}\AgdaSpace{}%
\AgdaSymbol{:}\AgdaSpace{}%
\AgdaSymbol{(}\AgdaFunction{𝓥}\AgdaSpace{}%
\AgdaOperator{\AgdaFunction{⇒}}\AgdaSpace{}%
\AgdaBound{𝒞}\AgdaSymbol{)}\AgdaSpace{}%
\AgdaBound{Γ}\AgdaSpace{}%
\AgdaBound{Δ}\AgdaSymbol{)}\AgdaSpace{}%
\AgdaSymbol{\{}\AgdaBound{i}\AgdaSymbol{\}}\AgdaSpace{}%
\AgdaSymbol{(}\AgdaBound{t}\AgdaSpace{}%
\AgdaSymbol{:}\AgdaSpace{}%
\AgdaBound{𝒞}\AgdaSpace{}%
\AgdaBound{Γ}\AgdaSpace{}%
\AgdaBound{i}\AgdaSymbol{)}\<%
\\
\>[12]\AgdaSymbol{→}\AgdaSpace{}%
\AgdaBound{𝑓}\AgdaSpace{}%
\AgdaSymbol{(}\AgdaField{map}\AgdaSpace{}%
\AgdaBound{δ}\AgdaSpace{}%
\AgdaBound{σ}\AgdaSpace{}%
\AgdaBound{t}\AgdaSymbol{)}\AgdaSpace{}%
\AgdaOperator{\AgdaDatatype{≡}}\AgdaSpace{}%
\AgdaField{map}\AgdaSpace{}%
\AgdaBound{δ'}\AgdaSpace{}%
\AgdaSymbol{(}\AgdaBound{𝑓}\AgdaSpace{}%
\AgdaOperator{\AgdaFunction{∘}}\AgdaSpace{}%
\AgdaBound{σ}\AgdaSymbol{)}\AgdaSpace{}%
\AgdaSymbol{(}\AgdaBound{𝑓}\AgdaSpace{}%
\AgdaBound{t}\AgdaSymbol{)}\<%
\\
\>[8]\AgdaFunction{Hmap}\AgdaSpace{}%
\AgdaBound{δ}\AgdaSpace{}%
\AgdaBound{δ'}\AgdaSpace{}%
\AgdaBound{nat}\AgdaSpace{}%
\AgdaSymbol{=}\AgdaSpace{}%
\AgdaField{Hpol}\AgdaSpace{}%
\AgdaBound{δ}\AgdaSpace{}%
\AgdaBound{δ'}\AgdaSpace{}%
\AgdaBound{nat}\AgdaSpace{}%
\AgdaSymbol{(λ}\AgdaSpace{}%
\AgdaBound{σ}\AgdaSpace{}%
\AgdaBound{v}\AgdaSpace{}%
\AgdaSymbol{→}\AgdaSpace{}%
\AgdaFunction{Hweaken}\AgdaSpace{}%
\AgdaBound{σ}\AgdaSpace{}%
\AgdaBound{v}\AgdaSymbol{)}\<%
\end{code}
And \AgdaFunction{subst}:
\begin{code}%
\>[8]\AgdaFunction{Hsubst}\AgdaSpace{}%
\AgdaSymbol{:}\AgdaSpace{}%
\AgdaSymbol{∀}\AgdaSpace{}%
\AgdaSymbol{\{}\AgdaBound{Γ}\AgdaSpace{}%
\AgdaBound{Δ}\AgdaSymbol{\}}\AgdaSpace{}%
\AgdaSymbol{(}\AgdaBound{σ}\AgdaSpace{}%
\AgdaSymbol{:}\AgdaSpace{}%
\AgdaSymbol{(}\AgdaFunction{𝓥}\AgdaSpace{}%
\AgdaOperator{\AgdaFunction{⇒}}\AgdaSpace{}%
\AgdaBound{𝒞}\AgdaSymbol{)}\AgdaSpace{}%
\AgdaBound{Γ}\AgdaSpace{}%
\AgdaBound{Δ}\AgdaSymbol{)}\<%
\\
\>[8][@{}l@{\AgdaIndent{0}}]%
\>[12]\AgdaSymbol{→}\AgdaSpace{}%
\AgdaSymbol{∀}\AgdaSpace{}%
\AgdaSymbol{\{}\AgdaBound{i}\AgdaSymbol{\}}\AgdaSpace{}%
\AgdaSymbol{(}\AgdaBound{t}\AgdaSpace{}%
\AgdaSymbol{:}\AgdaSpace{}%
\AgdaBound{𝒞}\AgdaSpace{}%
\AgdaBound{Γ}\AgdaSpace{}%
\AgdaBound{i}\AgdaSymbol{)}\<%
\\
\>[12]\AgdaSymbol{→}\AgdaSpace{}%
\AgdaBound{𝑓}\AgdaSpace{}%
\AgdaSymbol{(}\AgdaFunction{subst}\AgdaSpace{}%
\AgdaBound{σ}\AgdaSpace{}%
\AgdaBound{t}\AgdaSymbol{)}\AgdaSpace{}%
\AgdaOperator{\AgdaDatatype{≡}}\AgdaSpace{}%
\AgdaFunction{subst}\AgdaSpace{}%
\AgdaSymbol{(}\AgdaBound{𝑓}\AgdaSpace{}%
\AgdaOperator{\AgdaFunction{∘}}\AgdaSpace{}%
\AgdaBound{σ}\AgdaSymbol{)}\AgdaSpace{}%
\AgdaSymbol{(}\AgdaBound{𝑓}\AgdaSpace{}%
\AgdaBound{t}\AgdaSymbol{)}\<%
\\
\>[8]\AgdaFunction{Hsubst}\AgdaSpace{}%
\AgdaBound{σ}\AgdaSpace{}%
\AgdaBound{t}\AgdaSpace{}%
\AgdaSymbol{=}\AgdaSpace{}%
\AgdaFunction{Hmap}\AgdaSpace{}%
\AgdaFunction{id}\AgdaSpace{}%
\AgdaFunction{id}\AgdaSpace{}%
\AgdaSymbol{(λ}\AgdaSpace{}%
\AgdaBound{\AgdaUnderscore{}}\AgdaSpace{}%
\AgdaSymbol{→}\AgdaSpace{}%
\AgdaInductiveConstructor{refl}\AgdaSymbol{)}\AgdaSpace{}%
\AgdaBound{σ}\AgdaSpace{}%
\AgdaBound{t}\<%
\\
\\[\AgdaEmptyExtraSkip]%
\>[8]\AgdaOperator{\AgdaFunction{Hsubst𝕫/\AgdaUnderscore{}}}\AgdaSpace{}%
\AgdaSymbol{:}\AgdaSpace{}%
\AgdaSymbol{∀}\AgdaSpace{}%
\AgdaSymbol{\{}\AgdaBound{Γ}\AgdaSpace{}%
\AgdaBound{i}\AgdaSpace{}%
\AgdaBound{j}\AgdaSymbol{\}}\AgdaSpace{}%
\AgdaSymbol{(}\AgdaBound{t}\AgdaSpace{}%
\AgdaSymbol{:}\AgdaSpace{}%
\AgdaBound{𝒞}\AgdaSpace{}%
\AgdaBound{Γ}\AgdaSpace{}%
\AgdaBound{i}\AgdaSymbol{)}\AgdaSpace{}%
\AgdaSymbol{(}\AgdaBound{t'}\AgdaSpace{}%
\AgdaSymbol{:}\AgdaSpace{}%
\AgdaBound{𝒞}\AgdaSpace{}%
\AgdaSymbol{(}\AgdaBound{Γ}\AgdaSpace{}%
\AgdaOperator{\AgdaInductiveConstructor{◂}}\AgdaSpace{}%
\AgdaBound{i}\AgdaSymbol{)}\AgdaSpace{}%
\AgdaBound{j}\AgdaSymbol{)}\<%
\\
\>[8][@{}l@{\AgdaIndent{0}}]%
\>[12]\AgdaSymbol{→}\AgdaSpace{}%
\AgdaBound{𝑓}\AgdaSpace{}%
\AgdaSymbol{(}\AgdaFunction{subst}\AgdaSpace{}%
\AgdaSymbol{(}\AgdaOperator{\AgdaFunction{𝕫/}}\AgdaSpace{}%
\AgdaBound{t}\AgdaSymbol{)}\AgdaSpace{}%
\AgdaBound{t'}\AgdaSymbol{)}\AgdaSpace{}%
\AgdaOperator{\AgdaDatatype{≡}}\AgdaSpace{}%
\AgdaFunction{subst}\AgdaSpace{}%
\AgdaSymbol{(}\AgdaOperator{\AgdaFunction{𝕫/}}\AgdaSpace{}%
\AgdaBound{𝑓}\AgdaSpace{}%
\AgdaBound{t}\AgdaSymbol{)}\AgdaSpace{}%
\AgdaSymbol{(}\AgdaBound{𝑓}\AgdaSpace{}%
\AgdaBound{t'}\AgdaSymbol{)}\<%
\\
\>[8]\AgdaOperator{\AgdaFunction{Hsubst𝕫/\AgdaUnderscore{}}}\AgdaSpace{}%
\AgdaBound{t}\AgdaSpace{}%
\AgdaBound{t'}\AgdaSpace{}%
\AgdaKeyword{rewrite}\AgdaSpace{}%
\AgdaFunction{Hsubst}\AgdaSpace{}%
\AgdaSymbol{(}\AgdaOperator{\AgdaFunction{𝕫/}}\AgdaSpace{}%
\AgdaBound{t}\AgdaSymbol{)}\AgdaSpace{}%
\AgdaBound{t'}\AgdaSpace{}%
\AgdaSymbol{|}\AgdaSpace{}%
\AgdaOperator{\AgdaFunction{H𝕫/}}\AgdaSpace{}%
\AgdaBound{t}\AgdaSpace{}%
\AgdaSymbol{=}\AgdaSpace{}%
\AgdaInductiveConstructor{refl}\<%
\end{code}

Note that \AgdaFunction{Hmap} is a version of \AgdaField{Hnat}, but with
the \AgdaArgument{wk} argument removed. Also, we used the function
extensionality axiom here. The details are hidden.

\begin{code}%
\>[4]\AgdaKeyword{open}\AgdaSpace{}%
\AgdaModule{Hom}\AgdaSpace{}%
\AgdaSymbol{⦃...⦄}\AgdaSpace{}%
\AgdaKeyword{public}\<%
\end{code}

You can see an example of usage in the accompanying repository.

Apart from homomorphisms, we are also interested in the interactions of
substitution with \emph{itself}. This is the celebrated \textbf{substitution
lemma}:
$$t(x \mapsto s_1)(y \mapsto s_2) = t(y \mapsto s_2)(x\mapsto s_1(y \mapsto s_2))$$
under the condition that $x$ is not free in $s_2$. To prove this, we similarly
need to prove a version for renamings first. One would naturally ask whether
these two versions can be unified.

Here the coherence conditions get really nasty. The final goal is clear:
\begin{code}[hide]%
\>[4]\AgdaKeyword{record}\AgdaSpace{}%
\AgdaRecord{Stable}\AgdaSpace{}%
\AgdaSymbol{(}\AgdaBound{𝒞}\AgdaSpace{}%
\AgdaSymbol{:}\AgdaSpace{}%
\AgdaFunction{Scope}\AgdaSymbol{)}\AgdaSpace{}%
\AgdaSymbol{⦃}\AgdaSpace{}%
\AgdaBound{𝒞ˢ}\AgdaSpace{}%
\AgdaSymbol{:}\AgdaSpace{}%
\AgdaRecord{Syntax}\AgdaSpace{}%
\AgdaBound{𝒞}\AgdaSpace{}%
\AgdaSymbol{⦄}\AgdaSpace{}%
\AgdaSymbol{:}\AgdaSpace{}%
\AgdaPrimitive{Set₁}\AgdaSpace{}%
\AgdaKeyword{where}\<%
\\
\>[4][@{}l@{\AgdaIndent{0}}]%
\>[8]\AgdaKeyword{field}\<%
\\
\>[8][@{}l@{\AgdaIndent{0}}]%
\>[12]\AgdaField{map-comp}\AgdaSpace{}%
\AgdaSymbol{:}\AgdaSpace{}%
\AgdaSymbol{⦃}\AgdaSpace{}%
\AgdaBound{𝒜ʷ}\AgdaSpace{}%
\AgdaSymbol{:}\AgdaSpace{}%
\AgdaRecord{Weakening}\AgdaSpace{}%
\AgdaGeneralizable{𝒜}\AgdaSpace{}%
\AgdaSymbol{⦄}\<%
\\
\>[12][@{}l@{\AgdaIndent{0}}]%
\>[16]\AgdaSymbol{->}\AgdaSpace{}%
\AgdaSymbol{⦃}\AgdaSpace{}%
\AgdaBound{𝒟ˢ}\AgdaSpace{}%
\AgdaSymbol{:}\AgdaSpace{}%
\AgdaRecord{Syntax}\AgdaSpace{}%
\AgdaGeneralizable{𝒟}\AgdaSpace{}%
\AgdaSymbol{⦄}\<%
\\
\>[16]\AgdaSymbol{->}\AgdaSpace{}%
\AgdaSymbol{(}\AgdaBound{𝑔}\AgdaSpace{}%
\AgdaSymbol{:}\AgdaSpace{}%
\AgdaOperator{\AgdaFunction{[}}\AgdaSpace{}%
\AgdaGeneralizable{𝒟}\AgdaSpace{}%
\AgdaOperator{\AgdaFunction{⇒}}\AgdaSpace{}%
\AgdaBound{𝒞}\AgdaSpace{}%
\AgdaOperator{\AgdaFunction{]}}\AgdaSymbol{)}\<%
\\
\>[16]\AgdaSymbol{->}\AgdaSpace{}%
\AgdaSymbol{(}\AgdaBound{𝑓}\AgdaSpace{}%
\AgdaSymbol{:}\AgdaSpace{}%
\AgdaOperator{\AgdaFunction{[}}\AgdaSpace{}%
\AgdaGeneralizable{𝒜}\AgdaSpace{}%
\AgdaOperator{\AgdaFunction{⇒}}\AgdaSpace{}%
\AgdaGeneralizable{𝒟}\AgdaSpace{}%
\AgdaOperator{\AgdaFunction{]}}\AgdaSymbol{)}\<%
\\
\>[16]\AgdaSymbol{->}\AgdaSpace{}%
\AgdaSymbol{∀}\AgdaSpace{}%
\AgdaSymbol{\{}\AgdaBound{Γ}\AgdaSpace{}%
\AgdaBound{Δ}\AgdaSpace{}%
\AgdaBound{Θ}\AgdaSymbol{\}}\<%
\\
\>[16]\AgdaSymbol{->}\AgdaSpace{}%
\AgdaSymbol{(}\AgdaBound{σ}\AgdaSpace{}%
\AgdaSymbol{:}\AgdaSpace{}%
\AgdaSymbol{(}\AgdaFunction{𝓥}\AgdaSpace{}%
\AgdaOperator{\AgdaFunction{⇒}}\AgdaSpace{}%
\AgdaGeneralizable{𝒜}\AgdaSymbol{)}\AgdaSpace{}%
\AgdaBound{Γ}\AgdaSpace{}%
\AgdaBound{Δ}\AgdaSymbol{)}\AgdaSpace{}%
\AgdaSymbol{(}\AgdaBound{δ}\AgdaSpace{}%
\AgdaSymbol{:}\AgdaSpace{}%
\AgdaSymbol{(}\AgdaFunction{𝓥}\AgdaSpace{}%
\AgdaOperator{\AgdaFunction{⇒}}\AgdaSpace{}%
\AgdaGeneralizable{𝒟}\AgdaSymbol{)}\AgdaSpace{}%
\AgdaBound{Θ}\AgdaSpace{}%
\AgdaBound{Γ}\AgdaSymbol{)}\<%
\\
\>[16]\AgdaSymbol{->}\AgdaSpace{}%
\AgdaKeyword{let}\AgdaSpace{}%
\AgdaKeyword{instance}\AgdaSpace{}%
\AgdaBound{\AgdaUnderscore{}}\AgdaSpace{}%
\AgdaSymbol{=}\AgdaSpace{}%
\AgdaFunction{Syntaxʷ}\AgdaSpace{}%
\AgdaSymbol{⦃}\AgdaSpace{}%
\AgdaBound{𝒟ˢ}\AgdaSpace{}%
\AgdaSymbol{⦄}\AgdaSpace{}%
\AgdaKeyword{in}\<%
\\
\>[16]\AgdaSymbol{∀}%
\>[1662I]\AgdaSymbol{\{}\AgdaBound{i}\AgdaSymbol{\}}\AgdaSpace{}%
\AgdaSymbol{(}\AgdaBound{t}\AgdaSpace{}%
\AgdaSymbol{:}\AgdaSpace{}%
\AgdaBound{𝒞}\AgdaSpace{}%
\AgdaBound{Θ}\AgdaSpace{}%
\AgdaBound{i}\AgdaSymbol{)}\AgdaSpace{}%
\AgdaSymbol{->}\<%
\end{code}
\begin{code}%
\>[1662I][@{}l@{\AgdaIndent{1}}]%
\>[20]\AgdaField{map}\AgdaSpace{}%
\AgdaSymbol{(}\AgdaBound{𝑔}\AgdaSpace{}%
\AgdaOperator{\AgdaFunction{∘}}\AgdaSpace{}%
\AgdaBound{𝑓}\AgdaSymbol{)}\AgdaSpace{}%
\AgdaBound{σ}\AgdaSpace{}%
\AgdaSymbol{(}\AgdaField{map}\AgdaSpace{}%
\AgdaBound{𝑔}\AgdaSpace{}%
\AgdaBound{δ}\AgdaSpace{}%
\AgdaBound{t}\AgdaSymbol{)}\AgdaSpace{}%
\AgdaOperator{\AgdaDatatype{≡}}\AgdaSpace{}%
\AgdaField{map}\AgdaSpace{}%
\AgdaBound{𝑔}\AgdaSpace{}%
\AgdaSymbol{(}\AgdaField{map}\AgdaSpace{}%
\AgdaBound{𝑓}\AgdaSpace{}%
\AgdaBound{σ}\AgdaSpace{}%
\AgdaOperator{\AgdaFunction{∘}}\AgdaSpace{}%
\AgdaBound{δ}\AgdaSymbol{)}\AgdaSpace{}%
\AgdaBound{t}\<%
\end{code}

With suitably instantiated \AgdaArgument{𝑓} and
\AgdaArgument{𝑔}, this equation encapsulates both the renaming and substitution
lemmas.
But it is unclear what conditions should be imposed on \AgdaArgument{𝑓} and
\AgdaArgument{𝑔}. How should we proceed? Is this a dead end, and should we
now turn to more conventional methods of manipulating syntax with binding?
This is left as an \emph{exercise} for the reader.

\begin{quotation}
It is a question of foundations of mathematics, rather than mathematics itself;
or, at least, I hope so. The reply is left to the reader as an exercise.
(This phrase always means that the writer cannot do the problem himself.)
\attrib{\cite{difficult}, p. 15.}
\end{quotation}

\end{document}